\documentclass[lettersize,journal]{IEEEtran}

\usepackage[english]{babel}
\usepackage[utf8]{inputenc}
\usepackage[T1]{fontenc}
\usepackage{hyperref}

\usepackage{amsmath}
\usepackage{amsthm}
\usepackage{amsfonts}
\usepackage{mathtools} 
\usepackage{array}

\usepackage[linesnumbered, ruled, vlined]{algorithm2e}
\usepackage{multirow}
\usepackage{adjustbox} 
\usepackage{stfloats} 
\usepackage[caption=false]{subfig} 
\usepackage{booktabs} 
\usepackage{url}
\usepackage{verbatim}
\usepackage{cite}
\usepackage{wrapfig} 
\usepackage{graphicx}
\usepackage{textcomp}
\usepackage{pgfplots} 
\usepackage{tikz}
\usetikzlibrary{patterns} 
\usepgfplotslibrary{fillbetween} 
\usepackage{makecell}  

\usepackage[colorinlistoftodos]{todonotes} 
\usepackage{xcolor} 
\def\BibTeX{{\rm B\kern-.05em{\sc i\kern-.025em b}\kern-.08em 
    T\kern-.1667em\lower.7ex\hbox{E}\kern-.125emX}}
\usepackage{colortbl} 
\usepackage{pifont} 

\usepackage{eso-pic}
\AddToShipoutPictureBG*{%
  \put(575,700){\rotatebox{-90}{\parbox{21cm}{\centering\large\color{blue}
    \textit{This is the accepted version of an article published in \textbf{IEEE Transactions on Services Computing}. \\
    The final version is available at:} 
    \href{https://doi.org/10.1109/TSC.2025.3547235}{https://doi.org/10.1109/TSC.2025.3547235}}}}
}

\hypersetup{pdfborder=0 0 0}
\newcommand{\cmark}{\ding{51}}%
\newcommand\sbullet[1][.75]{\mathbin{\vcenter{\hbox{\scalebox{#1}{$\bullet$}}}}}
\newcommand{\highcell}{ \cellcolor{gray!45} } 
\newcommand{\medicell}{ \cellcolor{gray!25} }

\newcommand{\definitionname}{Def.}
\newcommand{\algorithmname}{Alg.}

\newcommand{\sectionname}{Sec.}
\definecolor{customblue}{HTML}{BDE0FF}
\definecolor{customgray}{HTML}{E6E6E6}
\definecolor{customred}{HTML}{FFCCCC}

\pgfplotsset{compat=1.13}

\theoremstyle{definition}
\newtheorem{defn}{Definition}[]

\makeatletter
\def\th@definition{
  \thm@notefont{}
  \normalfont 
}
\makeatother

\begin{document}

    \graphicspath{{figures/}}

\title{Efficient Online Computation of Business Process State From Trace Prefixes via N-Gram Indexing}

\author{

David Chapela-Campa,~\IEEEmembership{University of Tartu,~Estonia}, david.chapela@ut.ee

Marlon Dumas,~\IEEEmembership{University of Tartu,~Estonia}, marlon.dumas@ut.ee

}



\maketitle

\begin{abstract}
This paper addresses the following problem: Given a process model and an event log containing trace prefixes of ongoing cases of a process, map each case to its corresponding state (i.e., marking) in the model.
This state computation operation is a building block of other process mining operations, such as log animation and short-term simulation.
An approach to this state computation problem is to perform a token-based replay of each trace prefix against the model.
However, when a trace prefix does not strictly follow the behavior of the model, token replay may produce a state that is not reachable from the initial state of the process.
An alternative approach is to first compute an alignment between the trace prefix of each ongoing case and the model, and then replay the aligned trace prefix.
However, (prefix-)alignment is computationally expensive.
This paper proposes a method that, given a trace prefix of an ongoing case, computes its state in constant time on the length of the trace using an index that represents states as $n$-grams.
An empirical evaluation shows that the proposed approach has an accuracy comparable to that of the prefix-alignment approach, while achieving a throughput of hundreds of thousands of traces per second.
\end{abstract}



\section{Introduction\label{sec:introduction}}

\begin{figure}[t]
    \centering
        
    \begin{minipage}{0.99\columnwidth}
    	\centering
	    \subfloat[]{
            \includegraphics[width=\textwidth]{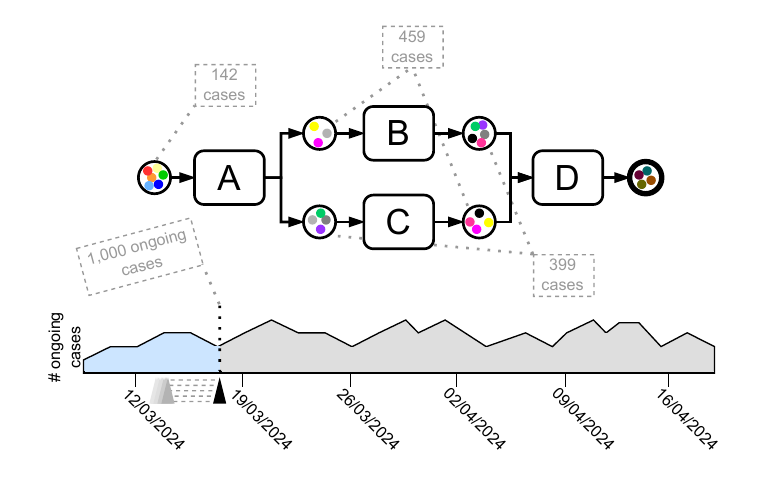}
	        \label{subfig:token-replay-example}
	    }
    \end{minipage}
    
    \begin{minipage}{0.99\columnwidth}
    	\centering
	    \subfloat[]{
            \includegraphics[width=\textwidth]{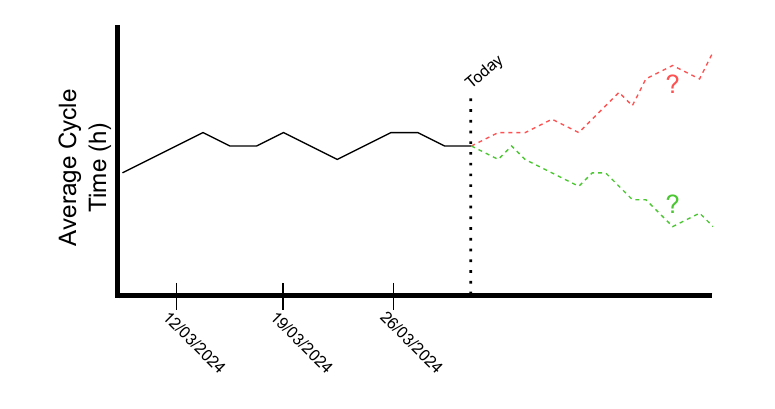}
	        \label{subfig:short-term-simulation-example}
	    }
    \end{minipage}

    \caption{Example of token-replay (a) and short-term simulation (b) scenarios.}
    \label{fig:motivation}
\end{figure}

\IEEEPARstart{P}{rocess Mining} (PM) is a set of techniques to discover, analyze, simulate, and monitor processes to optimize their performance, using event logs representing the execution of cases in a process~\cite{DBLP:books/sp/Aalst16}.
Several PM operations require us to determine the state of each ongoing case of a process, i.e., which activities may be executed next according to a process model.
The state of all ongoing cases is needed, for example, in log animation~\cite{DBLP:journals/sosym/LeoniSHA16}, to resume the visual replay of cases whenever the user advances the animation to an arbitrary point in the log's timeframe, e.g., by sliding the cursor, as illustrated in  \figurename~\ref{subfig:token-replay-example}. This figure illustrates the situation where the user has suddenly moved the cursor to a time point close to 19 March 2024. The log animator needs to react by updating the tokens in the process model to reflect the state of the process as of this time. Unless all the states of the process at every point in the log timeframe have been pre-computed and cached (which is computationally expensive for complex logs with long timeframes), this state computation operation needs to be performed efficiently to ensure interactive response times.

Similarly, in short-term (online) simulation~\cite{DBLP:conf/bpm/WynnDFHA07,reijers1999short,DBLP:journals/dke/RozinatWAHF09}, the state of each case needs to be determined to simulate the remainder activities of those cases, e.g., in order to start the simulation in the same state as the ongoing one, as illustrated in \figurename~\ref{subfig:short-term-simulation-example}.

\begin{figure}[t]
    \begin{minipage}{0.55\columnwidth}
    	\centering
	    \subfloat[]{
            \includegraphics[width=\textwidth]{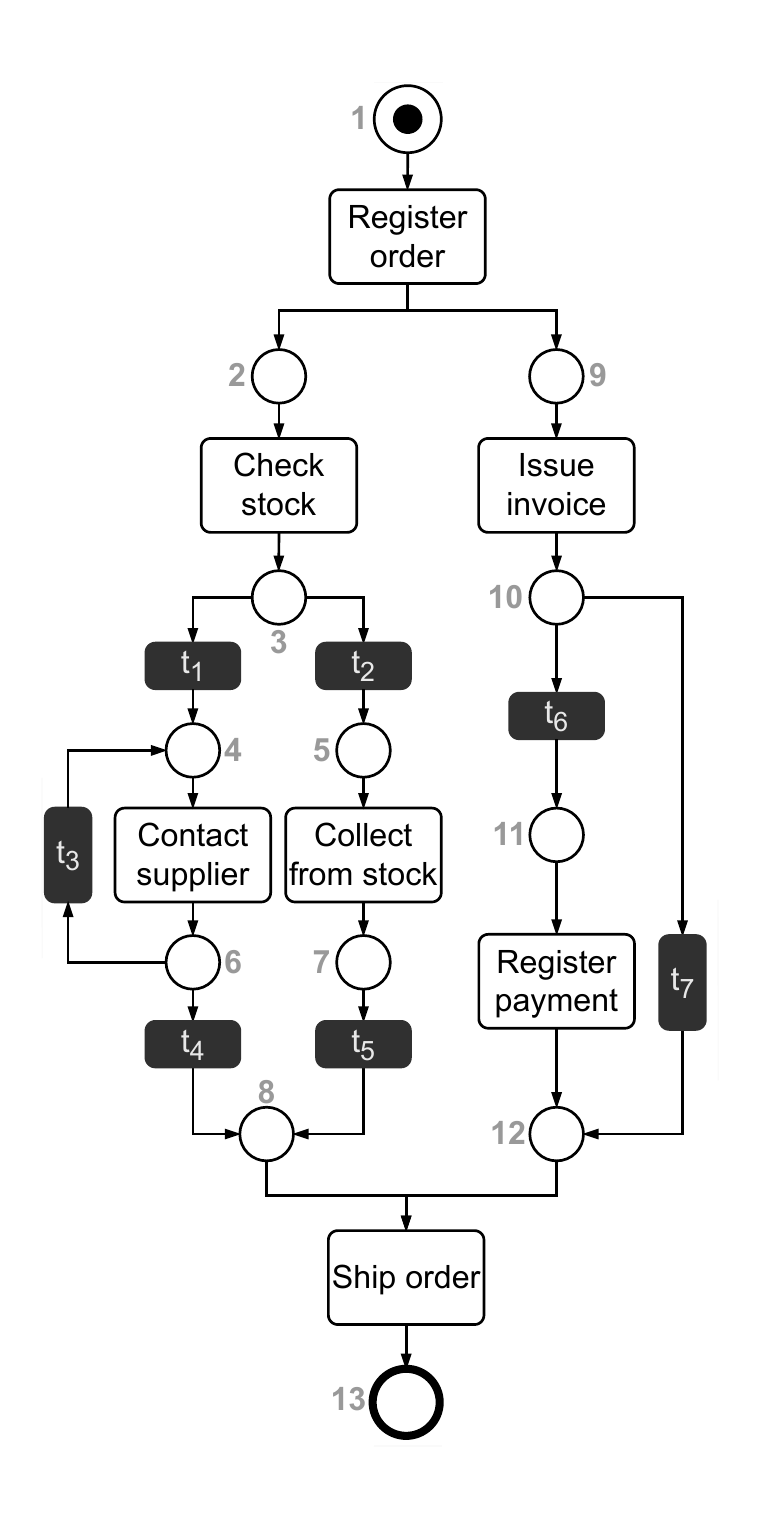}
	        \label{subfig:running-example}
	    }
    \end{minipage}
    \hfill
    \begin{minipage}{0.44\columnwidth}
    	\centering
	    \subfloat[]{
            \includegraphics[width=\textwidth]{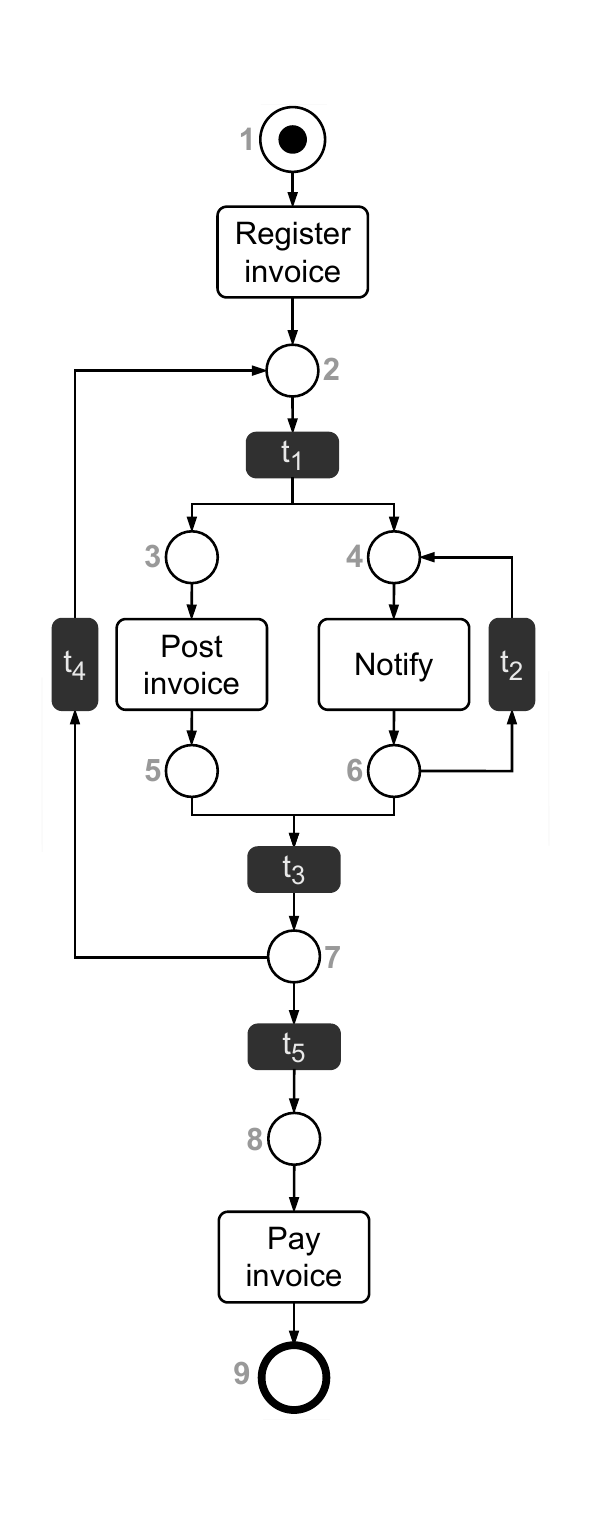}
	        \label{subfig:ambiguous-model}
	    }
    \end{minipage}

    \caption{Workflow nets of invoicing (a) and order handling (b) processes.}
    \label{fig:process-model-examples}
\end{figure}

In these settings, the problem of state computation may be defined as follows: Given a process model and an event log containing ongoing cases of a process, map each case to its corresponding state (i.e., marking) in the model.
For example, given the process model in \figurename~\ref{subfig:running-example}, the corresponding state of the ongoing case $\langle$Register order, Issue invoice, Check stock$\rangle$ is $\{3, 10\}$.
Note that the trace prefix of an ongoing case may not perfectly fit the model, or it may lead to multiple equally possible markings.
For example, given the process model in \figurename~\ref{subfig:ambiguous-model}, and the running case $\langle$Register invoice, Notify, Post invoice, Notify$\rangle$, both markings $\{5, 6\}$ and $\{3, 6\}$ are correct answers to the state computation problem.
The execution may loop back via either transition $t_{2}$ or transition $t_{4}$ to fire the second instance of ``Notify''.

This state computation problem can be approached by replaying the trace prefix on the process model~\cite{DBLP:conf/edoc/AdriansyahDA11,DBLP:journals/topnoc/BertiA21}.
However, in the case of non-fitting traces, i.e., traces with deviations from the behavior supported by the process model, (token-based) replay approaches produce markings that are not reachable from the start state.
Furthermore, in some circumstances, replay techniques may even create unreachable markings when replaying fitting traces (see \sectionname~\ref{sec:evaluation}).
This issue can be addressed by computing an optimal alignment between the trace prefix and the process model~\cite{DBLP:conf/bpm/SchusterZ20}, and then performing the replay using the resulting alignment.
However, both the prefix alignment and replay steps have exponential complexity.

In log animation or short-term simulation, there is a need to compute the state of thousands of ongoing cases in interactive times, which requires a method with low online complexity.

This paper proposes a method to compute the state of an ongoing case, given its trace prefix, in constant time on the length of the trace. 
The key idea is that the last $n$ executed activities ($n$ being a tunable parameter) are often sufficient to identify the current marking of an ongoing case.
Accordingly, the method builds the reachability graph of the process model and creates an index that maps the ending $n$-gram (i.e., last $n$ activities) of every possible trace prefix generated by the model, to the state(s) this prefix leads to.
The index is created offline.
At runtime, the state of an ongoing case is computed by searching for the ending $n$-gram in the index.

The paper reports on an evaluation of the proposed method to assess \textit{i)} its ability to accurately compute the state of ongoing cases (in both synthetic and real-life scenarios), and \textit{ii)} the efficiency of the proposed approach.
The evaluation relies on 20 synthetic event logs with different levels of noise and 12 event logs of real-life processes.

The rest of the paper is structured as follows.
\sectionname~\ref{sec:background} introduces background concepts on process mining and state computation.
\sectionname~\ref{sec:related-work} reviews related work.
\sectionname~\ref{sec:approach} describes the proposed method.
Finally, \sectionname~\ref{sec:evaluation} discusses the evaluation, and \sectionname~\ref{sec:conclusions} draws conclusions and sketches future work.

\section{Background\label{sec:background}}


We consider a process involving a set of \emph{activities} $\mathcal{A}$.
An \emph{event} $\varepsilon = (\varphi, \alpha, \tau)$ denotes an instance of activity $\alpha \in A$, where $\varphi$ identifies the process \emph{case} to which this event belongs, and $\tau$ denotes the timestamp of this event.\footnote{We work with a type of event log known as activity instance log, where there is one entry per activity instance and each activity instance has one timestamp (its end time).}
Accordingly, we write $\varphi(\varepsilon_{i})$, $\alpha(\varepsilon_{i})$, and $\tau(\varepsilon_{i})$ to denote, respectively, the case, the activity, and the timestamp associated with the event $\varepsilon_{i}$.
A \emph{trace} is a sequence of events $\Upsilon = \langle \varepsilon_{1}, \varepsilon_{2}, \ldots, \varepsilon_{q} \rangle$ with a common case identifier ($\varphi$).
We may represent a trace in simplified form as a sequence of activities sorted chronologically, e.g., $\langle$A, B, C$\rangle$.
An \emph{event log} is a collection of traces.
A \emph{(case) variant} is a unique activity sequence, and its \emph{frequency} is the number of cases whose corresponding traces match this activity sequence.
Finally, an \emph{ongoing case} is a process case that is currently in progress and has not yet reached its completion (e.g., when its execution in the process model leads to a state different than the final state).
The trace of an ongoing case is called its \emph{trace prefix}, and we call the sequence of its last $m$ consecutive activities its \emph{ending $m$-gram}.\footnote{
  There is no exact way to automatically assess whether a process case is terminated or not without the presence of a final activity in the process model.
  For example, if the process model contains a choice to loop back before its end, the replay of a terminated case results in a state previous to the loop choice.
  Nevertheless, this problem is orthogonal to that of finding the ongoing state of a trace prefix.}


At an abstract level, a \emph{process model} is a graphical representation of the set of traces of a process. 
There are many formalisms to represent process models, e.g., BPMN\footnote{\url{https://www.bpmn.org/}} or Petri nets~\cite{DBLP:conf/bpm/Aalst00}.
In this paper, we represent process models with \emph{labeled Petri nets}.

\begin{defn}[Labeled Petri net~\cite{DBLP:books/sp/Aalst16}\label{def:petri-net}]
    A (labeled) Petri net $PN = (P, T, F, A, l, c)$ consists of a set $P$ of places, a set $T$ of transitions, a set $F \subseteq (P \times T) \cup (T \times P)$ of flows that connect places with transitions, a set $A$ of activity labels, a function $l: T \rightarrow A$ that assigns a label to each transition in $T$, and a function $c: T \rightarrow \{silent, observable\}$ that assigns a class (i.e., silent or observable) to each transition in $T$.
\end{defn}

A flow $f = (p, t)$ is a directed edge that connects $p$ (i.e., its source) with $t$ (i.e., its target).
Given a Petri net $PN = (P, T, F, A, l, c)$, a place $p \in P$ is an \emph{input place} of a transition $t \in T$ iff $(p, t) \in F$, and an \emph{output place} of $t$ iff $(t, p) \in F$.
We use $\sbullet t$ to denote the set of input places of $t$, and $t \sbullet$ to denote the set of output places of $t$.
Similarly, $\sbullet p$ and $p \sbullet$ denote, respectively, the set of input and output transitions of $p$.


It must be noted that the structure defined by \definitionname~\ref{def:petri-net} 
allows for Petri nets with multiple connected elements (a disconnected graph) or infinite loops, among other structures.
Such structures are undesirable when modeling business processes, where the process workflow tends to have a well-defined start and end (e.g., ``Register invoice'' and ``Pay invoice'' in \figurename~\ref{subfig:ambiguous-model}).
Furthermore, the execution of each process instance follows predefined paths that must eventually lead to the end of the process~\cite{DBLP:books/sp/Aalst16}.
For this reason, we work with the notion of labeled Workflow net (WF-net).

\begin{defn}[Labeled Workflow net~\cite{DBLP:books/sp/Aalst16}\label{def:workflow-net}]
     A (labeled) Workflow net is a Petri net $W = (P, T, F, A, l, c)$ such that 
     \begin{itemize}
         \item $\exists!\ source \in P \mid \sbullet source = \emptyset$, i.e., there exists only one place with no incoming flows (i.e., $source$);
         \item $\exists!\ sink \in P \mid sink \sbullet = \emptyset$, i.e., there exists only one place with no outgoing flows (i.e., $sink$); and
         \item $\forall v \in \{P \cup T\}$, there exists a (potentially empty) directed path from $source$ to $v$, and from $v$ to $sink$.
     \end{itemize}
\end{defn}

\begin{figure}[t]
    \centering
    \includegraphics[width=0.95\columnwidth]{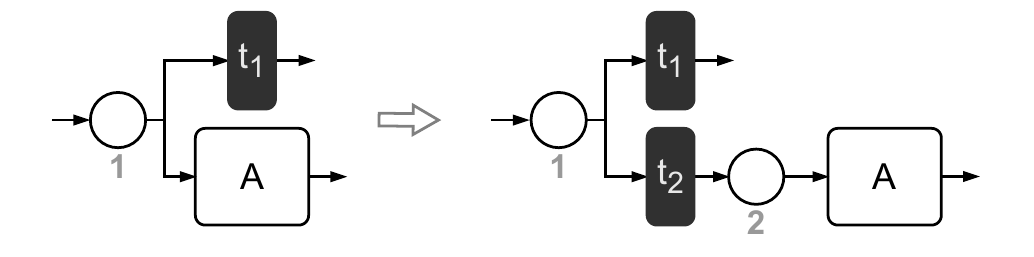}
    \caption{Transformation of an XOR-split place connected to both silent and observable transitions into an XOR-split connected to only silent transitions.}
    \label{fig:petri-net-transformation-MIX}
\end{figure}

\figurename~\ref{fig:process-model-examples} depicts two WF-nets.
In this representation, the places of a WF-net are depicted as circles, and the transitions as rectangles.
White-colored transitions are known as \emph{observable transitions}, and represent the activities of the process.
Conversely, black-colored transitions are known as \emph{silent transitions}, and capture the routing logic of the process (i.e., their execution does not map to any activity of the process).
Depending on the number of input and output transitions, a place $p$ can be classified as a sequential place ($\lvert \sbullet p \rvert = 1 \wedge \lvert p \sbullet \rvert = 1 $), an XOR-join ($\lvert \sbullet p \rvert > 1$), or an XOR-split ($\lvert p \sbullet \rvert > 1$).
Similarly, a transition $t$ can be classified as a sequential transition ($\lvert \sbullet t \rvert = 1 \wedge \lvert t \sbullet \rvert = 1 $), an AND-join ($\lvert \sbullet t \rvert > 1$), or an AND-split ($\lvert t \sbullet \rvert > 1$)~\cite{DBLP:books/sp/Aalst16}. 
Note that a place or transition can be classified as both join and split.
Without loss of generality, we assume that the output transitions of any XOR-split are all silent or all observable transitions.
An XOR-split with mixed silent and observable output transitions can be re-written by adding a silent transition followed by a place to the input of each output observable transition (see \figurename~\ref{fig:petri-net-transformation-MIX}).

Intuitively, an XOR-split is a \emph{decision point}: when the place is reached, one of the outgoing branches is activated (the mechanism for selection of a branch may be based on data conditions or probabilities). 
An AND-split is a \emph{parallel branching point}: when the transition is executed, all its outgoing branches are activated.
An XOR-join is a \emph{merge point}: as soon as one of the incoming branches completes, the outgoing branch is activated.
Finally, an AND-join is a \emph{synchronization point}: when all incoming branches have been completed, the outgoing branch is taken.


The semantics of WF-nets are defined as a token game~\cite{DBLP:books/sp/Aalst16}.
A \emph{marking} $M = \{ p \mid p \in P\}$ of $W$ is a set of places (i.e., having a token) that represents a state in the control-flow execution of the process.
Each $p \in M$ denotes the existence of a token in place $p$ (represented by a black dot in \figurename~\ref{fig:process-model-examples}).
Accordingly, the initial marking of a WF-net is $\{source\}$ and the final marking is $\{sink\}$.
Given a marking $M_{1}$, a transition $t \in T$ can be executed (i.e., it is said to be \textit{enabled}) iff $\sbullet t \subseteq M_{1}$, and its execution produces the marking $M_{2} = M_{1} \setminus \sbullet t \cup t \sbullet$.

Given a marking $M_{1}$, we use $M_{1} \xrightarrow{t} M_{2}$ to denote that $t$ is enabled in $M_{1}$, and its execution results in $M_{2}$.
We use $M_{1} \xrightarrow{\sigma} M_{2}$ to denote that executing the sequence of transitions $\sigma = \langle t_{1},t_{2},\ldots,t_{q} \rangle$ from marking $M_{1}$ leads to marking $M_{2}$.
A marking $M_{2}$ is said to be a \emph{reachable marking} from $M_{1}$ iff there exists a sequence (possibly empty) $\sigma$ of transitions such that $M_{1} \xrightarrow{\sigma} M_{2}$.
We use $M_{1} \xRightarrow{\sigma} M_{2}$ to denote that executing the sequence of silent transitions $\sigma = \langle t_{1},t_{2},\ldots,t_{q} \rangle$ such that $\forall t \in \sigma \mid c(t) = silent$, from marking $M_{1}$ leads to marking $M_{2}$.
Finally, a marking $M_{2}$ is said to be a reachable marking of a WF-net $W$ if it is reachable from $\{source\}$.

Although \definitionname~\ref{def:workflow-net} ensures the process model fulfills desirable structural properties, it still allows for structures that are undesired in well-designed business processes.
For example, the existence of deadlocks~\cite{iordache2006deadlock} or activities that can never be executed, among others~\cite{DBLP:books/sp/Aalst16}.
For this reason, in this paper, we focus on sound WF-nets.
In this context, a WF-net net $W = (P, T, F, A, l, c)$ with a source place $source$ and a sink place $sink$ is \emph{sound} iff 
\begin{itemize}
    \item $W$ is 1-safe, i.e., for all its reachable markings, places contain, at most, one token (a.k.a.\ \emph{safeness})\footnote{Due to the safeness property of sound WF-nets, a marking can be represented by a set.};
    \item the only reachable marking of $W$ containing $sink$ is $\{sink\}$, i.e., the final marking (a.k.a.\ \emph{proper completion});
    \item $\{sink\}$ is reachable for any reachable marking of $W$ (a.k.a.\ \emph{option to complete}); and
    \item for any $t \in T$, there is, at least, one reachable marking $M$ of $W$ such that $\sbullet t \subseteq M$ (a.k.a.\ \emph{absence of dead parts}).
\end{itemize}

\tablename~\ref{tab:token-replay-example} depicts an example of WF-net's semantics through the replay of the trace $\langle$Register order, Check stock, Issue invoice, Contact supplier, Register payment, Contact supplier, Ship order$\rangle$.
Note that, as mentioned above, the execution of the silent transitions (t$_1$, t$_6$, t$_3$, and t$_4$) does not correspond to observable behavior, and it is only triggered for routing purposes.

Interpreting the markings of a WF-net as its states, the execution of transitions moves the process execution from one state to another.
With this information, one can build an automaton capturing the behavior of the WF-net.
This is the \emph{reachability graph}~\cite{DBLP:books/sp/Aalst16}.

\begin{defn}[Reachability Graph\label{def:reachability-graph}]
    Given a WF-net $W = (P, T, F, A, l, c)$, a reachability graph of $W$ is a directed graph $RG = ( V, E, a )$ composed of a set $V \subseteq \mu$ of vertices, being $\mu$ the set of all reachable markings of $W$, a set $E$ of directed edges $e = (v_{s}, v_{t}) \mid v_{s}, v_{t} \in V$, and a mapping function $a : E \rightarrow T$ that maps each edge $e \in E$ with a transition $t \in T$, such that for every edge $e = (v_{s}, v_{t}) \in E$, $v_{s} \xrightarrow{a(e)} v_{t}$, i.e., the execution of the transition $a(e)$ in $W$, given the marking $v_{s}$, produces the marking $v_{t}$.
\end{defn}

Given a WF-net $W$, we define a reachability graph with $V$ being a subset or all reachable markings of $W$.
This implies that we allow for the existence of more than one reachability graph for $W$.
For example, we use \textit{complete reachability graph} to denote the reachability graph in which $V$ corresponds to the set of all reachable markings of $W$.
Thus, the reachability graph supports any path in $W$ from $source$ to $sink$.

\begin{table}[t]
    
    \centering
    \caption{
      Example of token-replay semantics over the BPMN model in \figurename~\ref{subfig:running-example}.
    }
    \label{tab:token-replay-example}

    \def\arraystretch{1.4}
    \begin{tabular}{l p{3cm} l}
        \toprule
        \multicolumn{1}{c}{\textbf{Current state}} & \multicolumn{1}{c}{\textbf{Enabled transitions}} & \multicolumn{1}{c}{\textbf{Firing transition}} \\ \midrule \midrule
        
        \{1\}      & \textsf{\{Register order\}}                                                   & \textsf{Register order}   \\
        \{2, 9\}   & \textsf{\{Check stock,} \newline \phantom{x}\textsf{Issue invoice\}}          & \textsf{Check stock}      \\
        \{3, 9\}   & \textsf{\{t$_{1}$, t$_{2}$, Issue invoice\}}                                  & \textsf{Issue invoice}    \\
        \{3, 10\}  & \textsf{\{t$_{1}$, t$_{2}$, t$_{6}$, t$_{7}$\}}                               & \textsf{t$_{1}$}          \\
        \{4, 10\}  & \textsf{\{Contact supplier, t$_{6}$, t$_{7}$\}}                               & \textsf{Contact supplier} \\
        \{6, 10\}  & \textsf{\{t$_{3}$, t$_{4}$, t$_{6}$, t$_{7}$\}}                               & \textsf{t$_{6}$}          \\
        \{6, 11\}  & \textsf{\{t$_{3}$, t$_{4}$,} \newline \phantom{x}\textsf{Register payment\}}  & \textsf{Register payment} \\
        \{6, 12\}  & \textsf{\{t$_{3}$, t$_{4}$\}}                                                 & \textsf{t$_{3}$}          \\
        \{4, 12\}  & \textsf{\{Contact supplier\}}                                                 & \textsf{Contact supplier} \\
        \{6, 12\}  & \textsf{\{t$_{3}$, t$_{4}$\}}                                                 & \textsf{t$_{4}$}          \\
        \{8, 12\}  & \textsf{\{Ship order\}}                                                       & \textsf{Ship order}       \\
        \{13\}     & $\emptyset$                                                                   & \textsf{-}                \\ \bottomrule
    \end{tabular}
\end{table}

\begin{defn}[Complete Reachability Graph\label{def:complete-reachability-graph}]
    Given a WF-net $W = (P, T, F, A, l, c)$, a reachability graph $RG = ( V, E, a )$ of $W$ is a \emph{complete reachability graph} of $W$ iff $V = \mu$, and for any $M_{1} \xrightarrow{t} M_{2} \mid t \in T \wedge M_{1}, M_{2} \in \mu$, there exists an edge $e = (v_{s}, v_{t}) \in E$ such that $a(e) = t$, $v_{s} = M_{1}$ and $v_{t} = M_{2}$. 
\end{defn}

As silent transitions only play a routing role, it is sometimes desirable to exclude them from the reachability graph.

\begin{defn}[Pure Reachability Graph\label{def:pure-reachability-graph}]
    Given a WF-net $W = (P, T, F, A, l, c)$, a pure reachability graph of $W$ is a directed graph $\mathit{PRG} = ( V, E, a )$ composed of a set $V \subseteq \mu$ of vertices, being $\mu$ the set of all reachable markings of $W$, a set $E$ of directed edges $e = (v_{s}, v_{t}) \mid v_{s}, v_{t} \in V$, and a mapping function $a : E \rightarrow T$ that maps each edge $e \in E$ with a transition $t \in T \mid c(t) = observable$, such that for every edge $e = (v_{s}, v_{t}) \in E$, $v_{s} \xrightarrow{\sigma_t} v_{t}$ where $\exists! t \in \sigma_t \mid c(t) = observable \wedge t = a(e)$ holds.
    This is, the execution of the transition $a(e)$ and a (potentially empty) set of silent transitions in $W$, given the marking $v_{s}$, produces the marking $v_{t}$.
\end{defn}

A pure reachability graph is a reachability graph where an edge $e = (v_{s}, v_{t}) \in E$ such that $a(e) = t$ does not represent that the execution of $t$ from marking $v_{s}$ leads to $v_{t}$, but that the execution of a sequence of transitions $\sigma_t$ (in which $t$ is the only observable transition) from marking $v_{s}$ leads to $v_{t}$.

\begin{defn}[Complete Pure Reachability Graph\label{def:complete-pure-reachability-graph}]
    Given a WF-net $W = (P, T, F, A, l, c)$, a pure reachability graph $\mathit{PRG} = ( V, E, a )$ of $W$ is a \emph{complete pure reachability graph} of $W$ iff for any $M_{1} \xrightarrow{t} M_{2} \mid M_{1}, M_{2} \in \mu \wedge c(t) = observable$, then there exists an edge $e = (v_{s}, v_{t}) \in E \mid a(e) = t \wedge v_{s} \xRightarrow{\sigma^{\prime}} M_{1} \wedge M_{2} \xRightarrow{\sigma^{\prime\prime}} v_{t}$.
\end{defn}

A (complete) pure reachability graph\footnote{
  Henceforth, we will use the term ``pure reachability graph'' to refer to the ``complete pure reachability graph''.
} adds the restriction that, for every path $M_{1} \xrightarrow{t} M_{2}$ in the WF-net such that $t$ is an observable transition (e.g., $\{5, 9\} \xrightarrow{\text{Collect from stock}} \{7, 9\}$ in \figurename~\ref{subfig:running-example}), there must be an edge in the reachability graph $(v_{s}, v_{t})$ such that its label is $t$, and one can go from $v_{s}$ to $M_{1}$ and from $M_{2}$ to $v_{t}$ firing only silent transitions. 

\begin{figure}[t]
    \centering
    
    \begin{minipage}{\columnwidth}
    	\centering
	    \subfloat[WF-net with AND-split and AND-join.]{
            \includegraphics[width=0.9\textwidth]{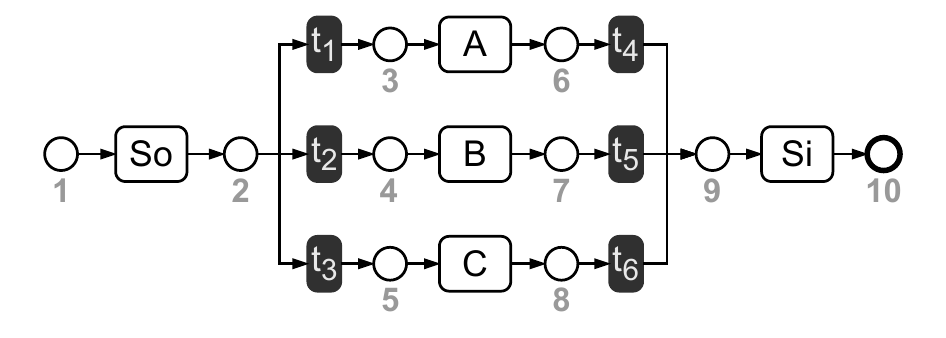}
	        \label{subfig:why-mixed-model}
	    }
    \end{minipage}
    
    \vspace{10pt}
    
    \begin{minipage}{\columnwidth}
        \begin{minipage}{0.495\textwidth}
            \centering
            \subfloat[Eager traversing.]{
                \includegraphics[width=\textwidth]{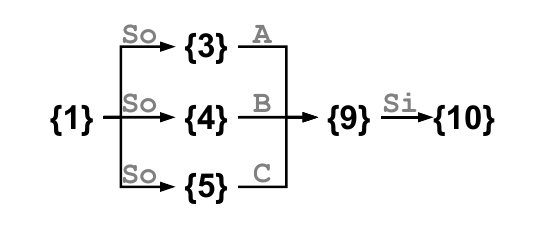}
                \label{subfig:why-mixed-eager}
            }
        \end{minipage}
        \hfill
        \begin{minipage}{0.495\textwidth}
            \centering
            \subfloat[Lazy traversing.]{
                \includegraphics[width=\textwidth]{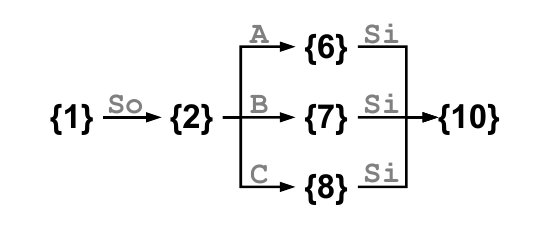}
                \label{subfig:why-mixed-lazy}
            }
        \end{minipage}
    \end{minipage}

    \vspace{10pt}
    
    \begin{minipage}{0.98\columnwidth}
        \centering
        \subfloat[Mixed traversing.]{
            \includegraphics[width=0.6\textwidth]{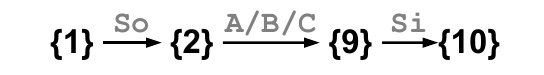}
            \label{subfig:why-mixed-mixed}
        }
    \end{minipage}
    
    \caption{Model with XOR-split and XOR-join structures (a) and three corresponding pure reachability graphs: eager-traversing (b), lazy-traversing (c), and lazy-splits/eager-joins traversing (d) policies.}
    \label{fig:why-mixed}

\end{figure}

By abstracting the execution of silent transitions, one can navigate through the pure reachability graph based only on the sequence of observable transitions (i.e., the activities of the process).
This creates a design choice regarding when to traverse silent activities: \textit{i)} in an \emph{eager} manner, i.e., as soon as they are enabled, or \textit{ii)} in a \emph{lazy} manner, only when it is strictly necessary to enable the observable transition that is going to be executed.
Different silent transition traversing policies yield different reachability graphs, all of them producing the same paths of observable transitions.

\figurename~\ref{subfig:why-mixed-model} shows a WF-net example with an XOR-split place ($2$) and an XOR-join place ($9$).
\figurename~\ref{subfig:why-mixed-eager} and \figurename~\ref{subfig:why-mixed-lazy} depict the pure reachability graphs of this WF-net following, respectively, eager and lazy traversing policies.
With an eager policy (\figurename~\ref{subfig:why-mixed-eager}), the execution of ``So'' advances the marking from $\{1\}$ to either $\{3\}$, $\{4\}$, or $\{5\}$, creating three different vertices in the reachability graph.
In this setting, the execution of ``A'', ``B'', or ``C'' leads to a single vertex, as the three of them eagerly traverse the silent transitions in the XOR-join ($9$).
Following a lazy policy creates a mirrored structure (\figurename~\ref{subfig:why-mixed-lazy}), where markings $\{3\}$, $\{4\}$, and $\{5\}$ get combined in $\{2\}$, but marking $\{9\}$ is split in three for the same reason.
As can be seen in \figurename~\ref{subfig:why-mixed-mixed}, lazily traversing XOR-splits and eagerly traversing all other silent transitions reduces the number of states in the reachability graph, while preserving the behavior.

\begin{figure*}[t]
    \centering
    
    \includegraphics[width=0.9\textwidth]{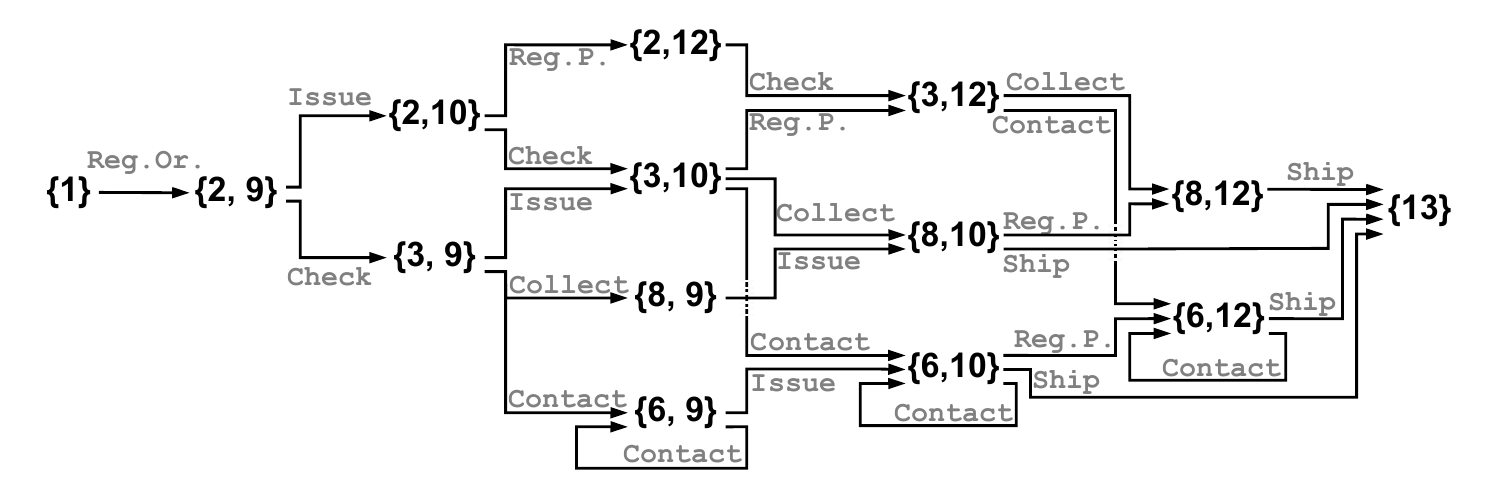}
    \caption{Pure reachability graph (lazy traversal policy only for XOR-splits) corresponding to the model in \figurename~\ref{subfig:running-example}.}
    \label{fig:reachability-graph-proposed}
\end{figure*}

\figurename~\ref{fig:reachability-graph-proposed} shows the pure reachability graph of the model depicted in \figurename~\ref{subfig:running-example} following a lazy traversal policy for XOR-splits, and eager for the rest of silent transitions.\footnote{
  Note that the chosen policy for AND transitions is irrelevant, as both eager and lazy policies would lead to one single vertex in the reachability graph (either with its input places or its output places).
}

\section{Related Work\label{sec:related-work}}

An intuitive and simple approach to compute the state of an ongoing case in a process model is to replay its trace prefix on the model, following the token game semantics sketched in \sectionname~\ref{sec:background}, and starting from the initial state -- i.e., $\{source\}$.
When the trace prefix does not fit the model, the (token-based) replay algorithm must apply heuristics that create and consume tokens to enable the execution of activities that are not enabled in the current state~\cite{DBLP:conf/edoc/AdriansyahDA11,DBLP:journals/topnoc/BertiA21}.
In some situations, this artificial creation and/or consumption of tokens may map the trace prefix to an unreachable state.
For example, given the process in \figurename~\ref{subfig:running-example}, the unfitting trace prefix $\langle$Register order, Check stock, Collect from stock, Register payment$\rangle$, may correspond to one of two (reachable) scenarios:
\textit{i)} the ongoing marking $\{8, 12\}$, where the invoice was issued but not registered due to an error in the system, thus, being the order ready to be shipped;
or \textit{ii)}, the ongoing marking $\{8, 9\}$, where the payment was wrongly registered due to the invoice not being generated, thus, the invoice still needs to be issued.
Nevertheless, the artificial creation of tokens to enable the last activity in the trace prefix, ``Register payment'', might generate the ongoing (unreachable) marking $\{8, 9, 12\}$. 

Furthermore, the adjustments made by replay techniques cannot ensure reachable markings even in the case of fitting traces.
For example, given the marking $\{5, 6\}$ for the process in \figurename~\ref{subfig:ambiguous-model}, the execution of ``Notify'' may loop back through transition t$_2$ and result in the marking $\{5, 6\}$, or loop back through transition t$_4$ and produce the marking $\{3, 6\}$.
In the context of fitting traces, only one of these decisions is correct, and it depends on the next executed activities.
As replay algorithms make this decision locally, based solely on the previous activities, even fitting trace prefixes may lead to unreachable wrong markings.

The impact of unreachable markings for the purpose of log animation, although undesirable, may be considered acceptable.
When the number of (artificially added) extra tokens in the animation is not high, they might be removed from the displayed animation once the trace is finished.
However, they still present an issue that can cause misinterpretations during the analysis of the animation.
Furthermore, unreachable markings are unsuitable for the purpose of short-term simulation.
Starting the simulation of an ongoing case from an unreachable marking might result in the creation of unfitting traces, or even cause deadlocks that block the execution of the case.

A possible approach to handle non-fitting trace prefixes is to use prefix-alignment techniques~\cite{DBLP:journals/ijdsa/ZelstBHDA19,DBLP:conf/bpm/SchusterZ20,DBLP:conf/caise/RaunNBA23} to align the prefix of the ongoing case with a prefix of the process model.
Prefix-alignment techniques are designed to compute the minimum error correction needed to transform a trace prefix (of an ongoing case) into a trace prefix that can be produced by the process model.
An error correction may be the skipping of an activity in the model that was not recorded in the case (move-on-model), or the skipping of an activity in the case that cannot be executed in the model (move-on-log).
Although prefix-aligment techniques are mainly designed for the purpose of conformance checking~\cite{DBLP:books/sp/CarmonaDSW18}, the resulting alignment can then be used to replay the trace prefix on the process model and compute its current state.
However, the problem of computing an optimal alignment has an exponential worst-case complexity.
There have been several studies on efficiently computing trace alignments, but the resulting algorithms either only work for certain classes of process models, or they strike a tradeoff between computational complexity and optimality of the trace alignment~\cite{DBLP:journals/tsc/SongXJZH17,DBLP:conf/caise/DongenCCT17,DBLP:conf/icpm/SyamsiyahD19}.

Other approaches have researched the mapping of trace prefixes to abstract states of a process.
For example, Burattin et al. represent in~\cite{DBLP:conf/bpm/BurattinZADC18} the state of a trace prefix as a multiset of observed sequence patterns, among a set of frequent patterns in the log; while the work of Lee et al.~\cite{DBLP:journals/is/LeeBMS21} maps a trace prefix to a state in a Hidden Markov Model.
Nevertheless, although these approaches allow us to estimate the fitness of a trace prefix w.r.t.\ a process model, they do not allow us to map a prefix to a state in the model.

Various approaches have been proposed in the field of concurrency theory to efficiently evaluate different types of queries over reachability graphs of Petri nets.
For example, Chen et al.~\cite{DBLP:journals/tods/ChenS21} presented a method to index a reachability graph in order to efficiently query whether there exists a path from a vertex $u$ to a vertex $v$ formed only by edges whose labels are in a given set (the Label-Constrained Reachability problem).
Meanwhile, Kashyap et al.~\cite{DBLP:conf/kbse/KashyapG05} propose to add metadata to the vertices of the graph, to efficiently calculate the reachability for certain types of predicates in the presence of concurrency.
Albeit tackling a different problem than the one we tackle in the present paper, these latter studies highlight the merits of indexing or enriching reachability graphs to efficiently tackle different types of queries. Within this body of work, our proposal can be seen as an indexing method to answer the question of what marking in the net a given trace leads to.

\begin{table}[t]
    {\centering
    \caption{
      Comparison of techniques for state computation.
    }
    \label{tab:related-work}

    \begin{tabular}{lccccc}
         \toprule
         & \multicolumn{1}{c}{\multirow{2}{*}{\begin{tabular}[c]{@{}c@{}}Petri\\ net\end{tabular}}} & \multicolumn{1}{c}{\multirow{2}{*}{\begin{tabular}[c]{@{}c@{}}Sound\\ WF-net\end{tabular}}} & \multicolumn{1}{c}{\multirow{2}{*}{\begin{tabular}[c]{@{}c@{}}Reach.\\ markings\end{tabular}}} & \multicolumn{2}{c}{Time complexity}                      \\ \cmidrule{5-6}
         & \multicolumn{1}{c}{}                                                                              & \multicolumn{1}{c}{}                                                                              & \multicolumn{1}{c}{}                                                                              & \multicolumn{1}{c}{Offline} & \multicolumn{1}{c}{Online} \\ \midrule \midrule
                   
        Token-based replay & \cmark                & \cmark                &                       & -                           & $\mathcal{O}(2^{n})$       \\
        Prefix-alignment   & \cmark                & \cmark                & \cmark                & -                           & $\mathcal{O}(2^{n})$       \\
        Our proposal       &                       & \cmark                & \cmark                & $\mathcal{O}(2^{n})$        & $\mathcal{O}(1)$           \\ \bottomrule
    \end{tabular}
    }        
    
    {\footnotesize * $\mathcal{O}(1)$ being constant time on the length of the case}
    
    {\footnotesize ** $\mathcal{O}(2^{n})$ being $n$ the number of transitions in the process model}
    
\end{table}

To summarize, \tablename~\ref{tab:related-work} compares the method proposed in this paper with token-based replay and prefix alignment approaches w.r.t.\ \textit{i)} their support for Petri nets and/or sound WF-nets, \textit{ii)} their ability to ensure the computation of reachable markings, and \textit{iii)} their time complexity.

\section{N-Gram Indexing\label{sec:approach}}

\begin{figure*}[b]
    \centering
    \includegraphics[width=\textwidth]{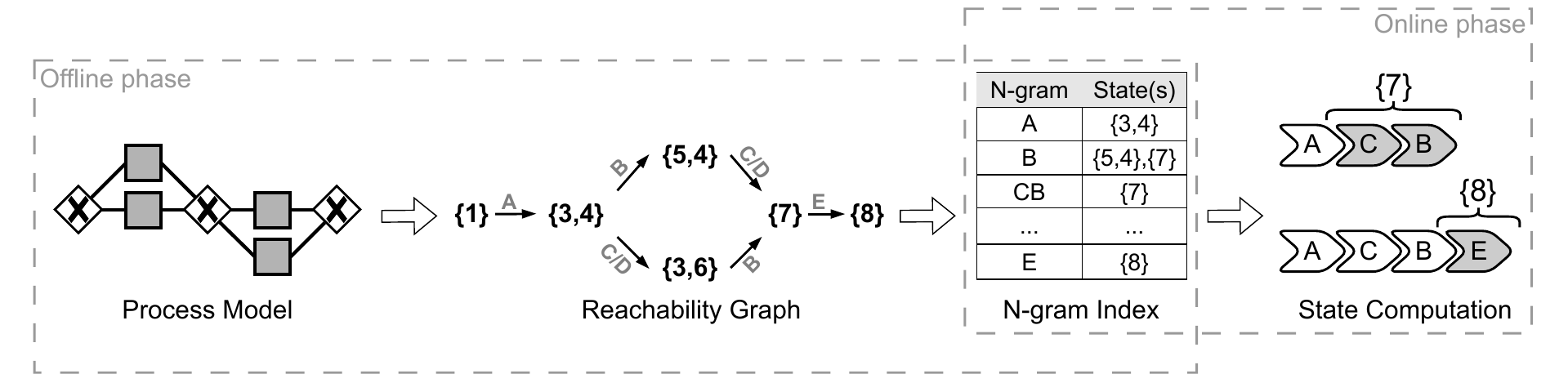}
    \caption{Overview of the approach proposed in this paper to compute the state of ongoing cases in constant time.}
    \label{fig:proposal-overview}
\end{figure*}

\figurename~\ref{fig:proposal-overview} depicts an overview of the approach proposed in this paper.
In an offline phase, we compute a pure reachability graph of the process model (\sectionname~\ref{subsec:reachability-graph}), which stores the possible states of a process case and how each activity displaces the execution from one state to another.
Then, we construct an index that associates each $m$-gram (i.e., sequence of $m$ consecutive activity labels), being $m \leq$ a predefined $n$, with the state(s) in the process model it leads to (\sectionname~\ref{subsec:n-gram-index}).
Thus, the state of an ongoing case can be computed in constant time on the length of the trace (linear on the selected $n$), at runtime, by searching in the index for the last $n$ activities of the prefix (\sectionname~\ref{subsec:compute-state}).

\subsection{Reachability Graph Generation\label{subsec:reachability-graph}}

The first step of our approach is to, given a process model, compute a (complete pure) reachability graph (see \definitionname~\ref{def:complete-pure-reachability-graph}) that models the behavior of the process as a state automaton.
When performing this computation, we want to follow a lazy silent transition traversing policy for XOR-split structures, while following an eager policy in all other cases.
One of the motivations for this lazy policy is to reduce the number of markings in the reachability graph (see \sectionname~\ref{sec:background}). 
In addition, when traversing a decision point, the selected path typically depends on execution data and/or on probabilities assigned to each path.
For example, in the process in \figurename~\ref{subfig:running-example}, the choice between transitions t$_1$ or t$_2$ is not made arbitrarily, but based on whether the ordered products are in stock or not.
In the context of log animation or short-term simulation, positioning the case in the state prior to a decision point allows the animation or simulation engine to select a branch based on branching probabilities or conditions, something that would not be possible if we returned a state after the decision point.

\begin{algorithm}[t]

\linespread{1.0}
\caption{Compute a pure reachability graph.}
\label{alg:reachability-graph}
{
    \SetAlgoLined
    \DontPrintSemicolon

    \SetKwFunction{AdvanceDecision}{advLazy}
    \SetKwFunction{AdvanceFully}{advEager}
    \SetKwFunction{Simulate}{replay}

    \SetKwProg{function}{Function}{}{}
    
    \KwIn{A WF-net $W = (P, T, F, A, l, c)$.}
    \KwOut{A pure reachability graph $\mathit{PRG} = ( V, E, a )$ of $W$ (lazy traversing policy for decision points).}

    $M_1 \gets$ \AdvanceDecision{$\{source\}$, $W$}\;\label{line:advance-dec-point}
    $V \gets \{M_1\}$\;\label{line:add-initial-as-vertex}
    $E \gets \emptyset$\;
    $M_{pend} \gets \{M_1\}$\;\label{line:save-initial-as-pending}
    $M_{expl} \gets \emptyset$\;
    \While{$M_{pend} \setminus M_{expl} \neq \emptyset$}{\label{line:advance-eager}
        $M_c \gets M_i \mid M_i \in M_{pend} \setminus M_{expl}$\;
        $M_{expl} \gets M_{expl} \cup \{M_c\}$\;
        \For{$(t_i, M_i) \in$ \AdvanceFully{$M_c$, $W$, $\emptyset$}}{
            $M_{j} \gets$ \Simulate{$t_i$, $M_i$, $W$}\;\label{line:execute-task}
            $M_{j}^{\prime} \gets$ \AdvanceDecision{$M_{j}$, $W$}\;\label{line:advance-dec-point-2}
            $V \gets V \cup \{M_{j}^{\prime}\}$\;\label{line:extend-graph-1}
            $e_j \gets (M_c, M_{j}^{\prime})$\;
            $E \gets E \cup \{e_j\}$\;
            $a(e_j) \gets l(t_i)$\;\label{line:extend-graph-2}
            $M_{pend} \gets M_{pend}\ \cup\ \{M_{j}^{\prime}\}$\;\label{line:add-pending}
        }
    }
    \Return{$(V, E, a)$}\;
}
\end{algorithm}

\begin{algorithm}[t]

\linespread{1.0}
\caption{Advance a marking through decision points.}
\label{alg:advance-decision-point}
{
    \SetAlgoLined
    \DontPrintSemicolon

    \SetKwFunction{Enabled}{isEnabled}
    \SetKwFunction{Rollback}{rollbk}

    \SetKwProg{function}{Function}{}{}
    
    \KwIn{A marking $M$, a WF-net $W = (P, T, F, A, l, c)$, and a set of explored markings $M_{expl}$.}
    \KwOut{A list of pairs $(t_i, M_i)$ where $t_i$ is an observable transition and $M_i$ is the marking that, advanced from $M$, enables $t_i$.}

    \function{\AdvanceFully{$M$, $W$, $M_{expl}$}}{
        $P \gets \emptyset$\;
        $M^{\prime} \gets $ \AdvanceDecision{$M$, $W$}\;
        $M_{pend} \gets \{M^{\prime}\}$\;
        \While{$M_{pend} \neq \emptyset$}{
            $M_{fut} \gets \emptyset$\;
            \For{$M_c \in M_{pend}$}{
                \If{$M_c \notin M_{expl}$}{
                    $M_{expl} \gets M_{expl} \cup \{M_c\}$\;
                    $T_s \gets \{t_i \in T \mid c(t_i) = silent\ \wedge $ \Enabled{$t_i$, $M_c$, $W$} $\}$\;\label{line:get-enabled-gateways}
                    \eIf{$T_s = \emptyset$}{\label{line:no-enabled-gateways}
                        $T_{enab} \gets \{t_i \in T \mid c(t_i) = observable\ \wedge $ \Enabled{$t_i$, $M_c$, $W$} $\}$\;
                        \For{$t_i \in T_{enab}$}{
                            $M_{r} \gets $ \Rollback{$M_c$, $M^{\prime}$, $t_i$, $W$} \;
                            $P \gets P \ \cup \ \{\langle t_i, M_{r} \rangle\}$\;
                        }
                    }{
                        \For{$t_s \in T_s$}{\label{line:get-enabled-gateway}
                            \eIf{$\lvert t_s \sbullet \rvert > 1$}{
                                $P \gets P \ \cup \ $\AdvanceFully{$M_c$, $W$, $M_{expl}$}\;
                            }{
                                $M_{c}^{\prime} \gets $ \Simulate{$t_i$, $M_c$, $W$}\;
                                $M_{fut} \gets M_{fut}\ \cup\ \{M_{c}^{\prime}$\}\;
                            }\label{line:replay-end}
                        }
                    }
                }
            }
            $M_{pend} \gets M_{fut}$\;
        }
        \KwRet $P$\;
    }
}
\end{algorithm}

\algorithmname~\ref{alg:reachability-graph} presents the pseudocode of the algorithm to compute such a reachability graph.
The algorithm starts by advancing the initial marking until reaching decision points or observable transitions (\algorithmname~\ref{alg:reachability-graph}:\ref{line:advance-dec-point}).
The function \texttt{advLazy()} takes a marking and a WF-net, and iteratively executes all enabled silent transitions that are not part of an XOR-split structure, advancing the marking in the model until all enabled elements are either observable transitions or decision points.
This marking, which represents the initial marking of the process, is added as a vertex to the reachability graph (\algorithmname~\ref{alg:reachability-graph}:\ref{line:add-initial-as-vertex}) and saved in the set of markings pending to explore (\algorithmname~\ref{alg:reachability-graph}:\ref{line:save-initial-as-pending}).
Then, the markings that are pending to explore -- i.e., the advanced initial markings and all markings that are added to the set in future iterations -- are processed individually.
This processing starts by advancing the current marking ($M_c$) through the enabled silent transitions of decision points, if any.
Such advancement, performed by the function \texttt{advEager()}, produces a list of pairs $\langle t_i, M_i \rangle$ with an observable transition ($t_i$) that can be enabled by advancing through silent transitions from $M_c$, and the marking result of such advancement ($M_i$).
Then, for each enabled observable transition $t_i$ and marking $M_i$, the algorithm replays the execution of $t_i$ (\algorithmname~\ref{alg:reachability-graph}:\ref{line:execute-task}), advances the resulting marking through non-XOR-split silent transitions (\algorithmname~\ref{alg:reachability-graph}:\ref{line:advance-dec-point-2}), saves the movement in the reachability graph (\algorithmname~\ref{alg:reachability-graph}:\ref{line:extend-graph-1}-\ref{line:extend-graph-2}), and adds the marking to the queue of markings pending to explore (\algorithmname~\ref{alg:reachability-graph}:\ref{line:add-pending}).

Importantly, the function \texttt{advEager()} does not trivially advance all branches until enabling each observable transition $t_i$, but only the necessary branches to enable it.
For example, following the model in \figurename~\ref{subfig:running-example}, for $M_c = \{3, 10\}$, a full advancement would produce the four markings $\{4, 11\}$, $\{4, 12\}$, $\{5, 11\}$, and $\{5, 12\}$.
Executing ``Register payment'' from both $\{4, 11\}$, and $\{5, 11\}$ would produce two markings ($\{4, 12\}$ and $\{5, 12\}$) in which the XOR-split of place $3$ is already traversed.
As previously reasoned in this section, the desired advanced marking in this case is $\{3, 11\}$.
To avoid this, the function \texttt{advEager()} (\algorithmname~\ref{alg:advance-decision-point}) first advances by replaying the execution of all enabled silent transitions (\algorithmname~\ref{alg:reachability-graph}:\ref{line:get-enabled-gateways},\ref{line:get-enabled-gateway}-\ref{line:replay-end}).
Then, when there are no more silent transitions enabled (\algorithmname~\ref{alg:reachability-graph}:\ref{line:no-enabled-gateways}), it goes over all enabled observable transitions and rollbacks the branches that were unnecessarily advanced to enable $t_i$.
For example, after advancing $\{3, 10\}$ to $\{4, 11\}$, both ``Contact supplier'' and ``Register payment'' are enabled.
For the pair where $t_i$ corresponds to ``Contact supplier'', the \texttt{rollbk()} function rollbacks the token in $11$ to $10$, as its advancement was not required to enable $t_i$ (producing $\{4, 10\}$).
Similarly, for the pair where $t_i$ corresponds to ``Register payment'', the rollbacked marking corresponds to $\{3, 11\}$.
For this, the function \texttt{rollbk()} needs to advance, one by one, all combinations of enabled flows in the powerset $\mathbb{P}(M_c)$, until finding the smaller one needed for $t_i$ to be enabled.
Then, it rollbacks the remaining flows to their corresponding ones in the original marking ($M^{\prime}$).

The size of the reachability graph is exponential on the number of transitions in the WF-net.
Accordingly, the reachability graph computation has an exponential time complexity.

\vspace{10pt}
We presented the algorithm to compute a reachability graph over WF-nets.
However, in light of the high use of BPMN models in the context of business process simulation~\cite{DBLP:journals/bise/RosenthalTS21} and business process managemet~\cite{DBLP:books/sp/DumasRMR18}, we analyzed the applicability of our proposal to Workflow graphs (WF-graphs), a subset of BPMN widely used for modeling business processes~\cite{DBLP:journals/is/FavreFV15}.
In this context, a marking would correspond to a set of flows in the net that ``contain a token'', the execution of tasks and gateways would move the tokens from (some of) their incoming flows to (some of) their outgoing flows~\cite{DBLP:journals/is/FavreFV15}, and the functions \texttt{advLazy()} and \texttt{advEager()} would advance through gateways instead of silent transitions.
Furthermore, an extension to support IOR gateways could be achieved by incorporating enablement and firing semantics of IOR gateways into the reachability graph computation algorithm, e.g., the semantics in~\cite{DBLP:conf/bpm/Volzer10}.

\subsection{N-gram Index Creation\label{subsec:n-gram-index}}

Given the reachability graph computed in the previous stage, we build an index structure that maps each sequence of $m$ consecutive activities ($m$-gram), such that $m \leq n$ that appear in any trace prefix of the model, to the marking(s) that this trace prefix leads to.

\begin{defn}[N-gram index]
    Given a positive integer number $n\in \mathbb{N}^{+}$, the \emph{$n$-gram index} of a complete pure reachability graph $\mathit{PRG} = ( V, E, a )$ is a list of key-value pairs $(key, value)$ such that, for all $e_1=(v_1,v_2), e_2=(v_2,v_3), \ldots, e_m=(v_m, v_{m+1}) \mid m \leq n$, and such that $e_1,e_2,\ldots,e_m \in E$, there is an entry $(key, value)$ such that $key = \langle a(e_1), a(e_2), \ldots, a(e_m) \rangle$ and $v_{m+1} \in value$, i.e., for all sequences of $n$ or less consecutive edges in $\mathit{PRG}$, the target marking of the last edge is part of the value associated to the sequence formed by the edges' labels.
\end{defn}

\algorithmname~\ref{alg:build-index} presents the pseudocode of an algorithm to build the $n$-gram index of a reachability graph.
The algorithm iterates over each marking $M_i$ in the reachability graph (\algorithmname~\ref{alg:build-index}:\ref{line:iterate-over-markings}) and retrieves the set of edge sequences ($E_{p}$), such as each edge sequence \textit{i)} forms a path in the reachability graph, \textit{ii)} is composed of $n$ or fewer edges, and \textit{iii)} the target of the last edge in the sequence is marking $M_i$ (\algorithmname~\ref{alg:build-index}:\ref{line:retrieve-edge-sequences}).
Then, for each of these edge sequences, the algorithm builds the $m$-gram as the sequence of the edges' labels (\algorithmname~\ref{alg:build-index}:\ref{line:build-key}), and associates the marking $M_i$ to such $m$-gram (\algorithmname~\ref{alg:build-index}:\ref{line:add-to-map-1},\ref{line:add-to-map-2}).

\begin{algorithm}[t]

\linespread{1.0}
\caption{Build an $n$-gram index.}
\label{alg:build-index}
{
    \SetAlgoLined
    \DontPrintSemicolon

    \SetKwProg{function}{Function}{}{}
    
    \KwIn{A (pure) reachability graph $\mathit{PRG} = ( V, E, a )$ and a positive integer $n$.}
    \KwOut{The $n$-gram index of $\mathit{PRG}$ as a function $\Pi$.}

    \For{$M_i \in V$}{\label{line:iterate-over-markings}
        $E_{p} \gets \{\langle (v_1,v_2), (v_2,v_3), \ldots, (v_m, v_{m+1}) \rangle \mid m \leq n \wedge v_{m+1} = M_i \}$\;\label{line:retrieve-edge-sequences}
        \For{$E_i = \langle e_1, e_2, \ldots, e_m \rangle \in E_{p}$}{
            $key \gets \langle a(e_1), a(e_2), \ldots, a(e_m) \rangle$\;\label{line:build-key}
            \eIf{$key \in \Pi$}{
                $\Pi(key) \gets \Pi(key) \cup \{M_i\}$\;\label{line:add-to-map-1}
            }{
                $\Pi(key) \gets \{M_i\}$\;\label{line:add-to-map-2}
            }
        }
    }
    \Return{$\Pi$}\;
}
\end{algorithm}

\tablename~\ref{tab:n-gram-index-example} depicts the $n$-gram index (for $n = 3$) corresponding to the reachability graph in \figurename~\ref{fig:reachability-graph-proposed}.
It must be noted that the $n$-gram index behaves as a monotonic function when the $n$-gram grows backward.
For example, in \figurename~\ref{fig:reachability-graph-proposed}, the $1$-gram $\langle$Check stock$\rangle$ leads to states $\{3, 9\}$, $\{3, 10\}$, and $\{3, 12\}$.
Any $m$-gram $\mid m > 1$ ending in ``Check stock'' must lead to a subset of those markings.
Thus, when building the $n$-gram index, the backward expansion of a growing $m$-gram can be stopped when reaching a deterministic solution.
For example, in \tablename~\ref{tab:n-gram-index-example}, the $1$-gram $\langle$Ship order$\rangle$ leads to one single state (deterministic).
Thus, it is unnecessary to store any $2$-gram with ``Ship order'' as its last activity.
Conversely, the $1$-gram $\langle$Contact supplier$\rangle$ leads to 3 different markings and, thus, it is expanded until all the resulting $m$-grams are deterministic, or $m = 3$ (the maximum size for this example).

\begin{table}[t]
    \centering
    \caption{
      $3$-gram index of the reachability graph in \figurename~\ref{fig:reachability-graph-proposed}.
    }
    \label{tab:n-gram-index-example}
    
    \def\arraystretch{1.18}
    \begin{tabular}{p{5cm} p{3cm}}
        \toprule
        \multicolumn{1}{c}{\textbf{$n$-gram}}                              & \multicolumn{1}{c}{\textbf{State(s)}}   \\ \midrule \midrule 
        
        $\langle$ Register order $\rangle$                                    & \{2, 9\}                                    \\
        $\langle$ Check stock $\rangle$                                       & \{3, 9\}, \{3, 10\}, \{3, 12\}              \\
        $\langle$ Contact supplier $\rangle$                                  & \{6, 9\}, \{6, 10\}, \{6, 12\}              \\
        $\langle$ Collect from stock $\rangle$                                & \{8, 9\}, \{8, 10\}, \{8, 12\}              \\
        $\langle$ Issue invoice $\rangle$                                     & \{2, 10\}, \{3, 10\},  \newline \phantom{xx} \{6, 10\}, \{8, 10\}  \\
        $\langle$ Register payment $\rangle$                                  & \{2, 12\}, \{3, 12\},  \newline \phantom{xx} \{6, 12\}, \{8, 12\}  \\
        $\langle$ Ship order $\rangle$                                        & \{13\}                                      \\
        
        $\langle$ Register order, Check stock $\rangle$                       & \{3, 9\}                                    \\
        $\langle$ Issue invoice, Check stock $\rangle$                        & \{3, 10\}                                   \\
        $\langle$ Register payment, Check stock $\rangle$                     & \{3, 12\}                                   \\
        
        $\langle$ Check stock, Contact supplier $\rangle$                     & \{6, 9\}, \{6, 10\}, \{6, 12\}              \\
        $\langle$ Contact supplier, Contact supplier $\rangle$                & \{6, 9\}, \{6, 10\}, \{6, 12\}              \\
        $\langle$ Issue invoice, Contact supplier $\rangle$                   & \{6, 10\}                                   \\
        $\langle$ Register payment, Contact supplier $\rangle$                & \{6, 12\}                                   \\
        
        $\langle$ Check stock, Collect from stock $\rangle$                   & \{8, 9\}, \{8, 10\}, \{8, 12\}              \\
        $\langle$ Issue invoice, Collect from stock $\rangle$                 & \{8, 10\}                                   \\
        $\langle$ Register payment, Collect from stock $\rangle$              & \{8, 12\}                                   \\
        
        $\langle$ Register order, Issue invoice $\rangle$                     & \{2, 10\}                                   \\
        $\langle$ Check stock, Issue invoice $\rangle$                        & \{3, 10\}                                   \\
        $\langle$ Collect from stock, Issue invoice $\rangle$                 & \{8, 10\}                                   \\
        $\langle$ Contact supplier, Issue invoice $\rangle$                   & \{6, 10\}                                   \\
        
        $\langle$ Register order, Register payment $\rangle$                  & \{2, 10\}                                   \\
        $\langle$ Check stock, Register payment $\rangle$                     & \{3, 10\}                                   \\
        $\langle$ Collect from stock, Register payment $\rangle$              & \{8, 10\}                                   \\
        $\langle$ Contact supplier, Register payment $\rangle$                & \{6, 10\}                                   \\

        $\langle$ Check stock, Contact supplier,  \newline \phantom{xxxxxxxxxxxxxxxx} Contact supplier $\rangle$        & \{6, 9\}, \{6, 10\}, \{6, 12\}  \\
        $\langle$ Contact supplier, Contact supplier,  \newline \phantom{xxxxxxxxxxxxxxxx} Contact supplier $\rangle$   & \{6, 9\}, \{6, 10\}, \{6, 12\}  \\
        $\langle$ Issue invoice, Contact supplier,  \newline \phantom{xxxxxxxxxxxxxxxx} Contact supplier $\rangle$      & \{6, 10\}                       \\
        $\langle$ Register payment, Contact supplier,  \newline \phantom{xxxxxxxxxxxxxxxx} Contact supplier $\rangle$   & \{6, 12\}                       \\
        
        $\langle$ Register order, Check stock,  \newline \phantom{xxxxxxxxxxxxxxxx} Contact supplier $\rangle$          & \{6, 9\}                        \\
        $\langle$ Issue invoice, Check stock,  \newline \phantom{xxxxxxxxxxxxxxxx} Contact supplier $\rangle$           & \{6, 10\}                       \\
        $\langle$ Register payment, Check stock,  \newline \phantom{xxxxxxxxxxxxxxxx} Contact supplier $\rangle$        & \{6, 12\}                       \\
        
        $\langle$ Register order, Check stock,  \newline \phantom{xxxxxxxxxxxxxxxx} Collect from stock $\rangle$        & \{8, 9\}                        \\
        $\langle$ Issue invoice, Check stock,  \newline \phantom{xxxxxxxxxxxxxxxx} Collect from stock $\rangle$         & \{8, 10\}                       \\
        $\langle$ Register payment, Check stock,  \newline \phantom{xxxxxxxxxxxxxxxx} Collect from stock $\rangle$      & \{8, 12\}                       \\ \bottomrule
    \end{tabular}
\end{table}

Given the monotonicity of the $n$-gram index, there exist (pure) reachability graphs -- and, by extension, WF-nets -- for which the $n$-gram index would never need to grow the prefixes more than a certain $m$, $m < n$.
We call this the \emph{$K$-complexity} of a WF-net, and it corresponds to the maximum $n$-gram size needed to unambiguously identify the ongoing state of a fitting case.
The $K$-complexity of a sequential WF-net is clearly $K=1$.
However, when a $p$-branch parallel structure is present, the $K$-complexity of the model equals to the sum of the lengths of the $p-1$ longest branches, plus one.
In this context, the length of a branch is equal to the number of activities of its longest sequence.
Finally, one can trivially see that the $K$-complexity of a WF-net is infinite if there is a loop in a parallel branch (e.g., the running example of \figurename~\ref{subfig:running-example}).

\subsection{Online State Computation\label{subsec:compute-state}}

The final stage of our proposal is to, given the trace prefix of an ongoing case of the process, compute its state in the process model.

\algorithmname~\ref{alg:compute-ongoing-state} presents the pseudocode of this stage.
Given the trace prefix of an ongoing case, we first filter out the activities that are not represented in the process model (\algorithmname~\ref{alg:compute-ongoing-state}:\ref{line:filter-ongoing-case}).
If the filtered trace prefix is empty, the ongoing state corresponds to the initial marking of the WF-net, i.e., $\{source\}$ (\algorithmname~\ref{alg:compute-ongoing-state}:\ref{line:assign-initial-marking}).\footnote{
  Note that the filtering out of the activities not represented in the model is equivalent to the ``\textit{move-on-log}'' performed in alignment techniques.
}
Otherwise, to compute the state of the filtered trace prefix, we search the index for each of its ending $m$-grams ($m \leq n$).
We will always find a match, since the ending $1$-gram (a single activity) is a $1$-gram of the reachability graph and hence included in the index (\algorithmname~\ref{alg:compute-ongoing-state}:\ref{line:assign-1-gram}).
We retain the matching index entry with the smallest $m$ that is associated with one single state, i.e., no ambiguity, or else the index entry with the smallest number of associated states, i.e., the least ambiguous (\algorithmname~\ref{alg:compute-ongoing-state}:\ref{line:assign-m-gram}).
In the latter scenario, we randomly select one of the states (\algorithmname~\ref{alg:compute-ongoing-state}:\ref{line:return-random}).

Following with the example in \tablename~\ref{tab:n-gram-index-example}, and given the trace $\langle$Register order, Issue invoice, Check stock, Collect from stock$\rangle$, the first search would be for the $1$-gram $\langle$Collect from stock$\rangle$.
As the result of this search is a set of three states, the search would continue with the $2$-gram $\langle$Check stock, Collect from stock$\rangle$, which also returns a set of three states.
Finally, the search would grow the $m$-gram to $\langle$Issue invoice, Check stock, Collect from stock$\rangle$, which returns the state $\{8, 10\}$ as a result.
Thus, when resuming the simulation of this ongoing case, two tokens would be placed in places 8 and 10 of \figurename~\ref{subfig:running-example}, which enables the execution of ``Register payment'' (through t$_6$) and ``Ship order'' (through t$_7$).

The worst-case complexity of one search in the $n$-gram index is $\mathcal{O}(\gamma)$, being $\gamma$ the number of $m$-grams in the index.
However, if implemented as a hash table, its complexity is $\mathcal{O}(1)$ amortized~\cite{cormen2022introduction,sedgewick2011algorithms}.
In the worst-case scenario (non-deterministic solution), our proposal needs to perform $n$ searches in the $n$-gram index.
Thus, the complexity of this online step is $\mathcal{O}(n)$.

\begin{algorithm}[t]

\linespread{1.0}
\caption{Compute the state of an ongoing case.}
\label{alg:compute-ongoing-state}
{
    \SetAlgoLined
    \DontPrintSemicolon

    \SetKwProg{function}{Function}{}{}
    
    \KwIn{An $n$-gram index modeled as a function $\Pi$, a WF-net $W = (P, T, F, A, l, c)$ of the process, and a trace $\Upsilon = \langle \varepsilon_{1}, \varepsilon_{2}, \ldots, \varepsilon_{q} \rangle$ of an ongoing case.}
    \KwOut{The ongoing state of $\Upsilon$.}

    $\Upsilon^{\prime} \gets \langle \varepsilon_{i} \mid \varepsilon_i \in \Upsilon \wedge \alpha(\varepsilon_i) \in A \wedge c(\varepsilon_i) = observable \rangle$\;\label{line:filter-ongoing-case}
    \eIf{$\lvert \Upsilon^{\prime} \rvert = 0$}{
        $M \gets \{\{source\}\}$\;\label{line:assign-initial-marking}
    }{
        $key = \langle \alpha(\varepsilon_r) \mid r = \lvert \Upsilon^{\prime} \rvert \rangle$\;
        $M_c \gets \Pi(key)$\;\label{line:assign-1-gram}
        $m \gets 1$\;
        \While{$\lvert M_c \rvert > 1 \wedge m < n$}{
            $key = \langle \alpha(\varepsilon_{q-i}) \mid \varepsilon_{q-i} \in \Upsilon \wedge 0\leq i \leq m \rangle$\;
            $M_c \gets \Pi(key)$\;\label{line:assign-m-gram}
        }
    }
    \Return{$M \mid M \in M_c$}\;\label{line:return-random}
}
\end{algorithm}

\section{Evaluation\label{sec:evaluation}}

This section reports on an experimental evaluation of: \textit{i)} the impact of noise on the ability of the proposed approach to accurately compute the state of ongoing cases, \textit{ii)} its accuracy in real-life scenarios where the state of the process is unknown, and \textit{iii)} its efficiency.
The first part of the evaluation addresses the following question:

\begin{description}
\item
    \textbf{Evaluation Question 1 (EQ1)}: \textit{is the proposed approach able to accurately compute the state of an ongoing case under different levels of noise?}
\end{description}

To assess this, we designed a set of simulation scenarios covering different levels of complexity.
We applied different levels of noise to the designed scenarios in order to study the impact of possible deviations from the expected behavior, as well as of wrongly recorded data.

The second part of the evaluation addresses the accuracy on real-life processes, where cases often deviate from the expected behavior and the actual state of a case is unknown.
This part addresses the following question: 

\begin{description}
\item
    \textbf{Evaluation Question 2 (EQ2)}: \textit{given the ongoing case of a real-life process, where the state is unknown, is the proposed approach able to compute a future-equivalent state?}
\end{description}

The state of an ongoing case in real-life processes is typically unknown, as the only available information is the recorded activity instances and their attributes.
Thus, we seek to compute future-equivalent states, where two states are future-equivalent if they allow for the execution of the same remaining behavior.

This part of the evaluation is also designed to validate the efficiency of the proposed approach in real-life scenarios, by addressing the following question:

\begin{description}
\item \textbf{Evaluation Question 3 (EQ3)}: \textit{is the approach able to handle thousands of traces per second?}
\end{description}

\subsection{Synthetic Evaluation}

This section describes the evaluation performed to validate the accuracy of the proposal when estimating the state of ongoing cases of a process under different levels of noise (EQ1).

\medskip
\noindent\textbf{Datasets \& Setup.}
To evaluate EQ1, we designed five simulation process models with different complexities.
One sequential process with two decision points (``Seq''), and one variant of this process with loops (``Loop'').
Both processes present a $K$-complexity (see \sectionname~\ref{subsec:n-gram-index}) of one, meaning that, in a perfect-fit case, the last executed activity is enough to denote the ongoing state.
In addition, three more processes containing parallel structures of two (``K3''), three (``K5''), and five (``K10'') concurrent branches with a $K$-complexity of three, five, and ten, respectively.
We generated a simulated log of 1000 cases for each of these models, and retained the first $m$ recorded activities of each case to represent ongoing cases.
For each case, $m$ is a random number between three (minimum required size for the noise injection commented below) and the number of events recorded in that case.
Finally, for each event log (``Raw''), we injected three levels of noise by randomly applying, to each ongoing case, one (``Noise-1''), two (``Noise-2''), and three (``Noise-3'') operations from \textit{i)} adding a new event of a random activity at a random position, \textit{ii)} deleting a random event, or \textit{iii)} swapping the order of two random consecutive events.

As a baseline, we used the token-based replay approach presented in~\cite{DBLP:journals/topnoc/BertiA21}.
This approach ($\mathit{TokenR}$) follows the token-replay semantics described in \sectionname~\ref{sec:background}.
In non-conforming situations, where the next recorded activity is not enabled given the current marking, the replay algorithm applies a set of heuristics to artificially add tokens in order to enable it.\footnote{We used the latest implementation available in PM4PY~\cite{DBLP:journals/simpa/BertiZS23}, a well-known Python library providing a varied set of tools for process mining.}
As commented in \sectionname~\ref{sec:related-work}, token-replay techniques might produce unreachable states in unfitting situations.
For this reason, we considered an approach based on prefix-alignments~\cite{DBLP:conf/bpm/SchusterZ20} as a second baseline ($\mathit{PrefAl}$).
We used the prefix-alignment proposal presented in~\cite{DBLP:conf/bpm/SchusterZ20} in order to obtain a fitting alignment and the result marking of each ongoing case.
We selected the approach proposed in~\cite{DBLP:conf/bpm/SchusterZ20}, as it includes a prefix-caching strategy to avoid recalculating prefix-alignments for event sequences that were already observed in the past, thus reducing computational time.

Although the state of an ongoing case in a simulated scenario is known, the injection of noise compromises its validity.
For example, given the WF-net in \figurename~\ref{subfig:running-example} and the prefix with noise $\langle$Register order, Register payment, Issue invoice$\rangle$, both states $\{2, 10\}$ (considering the noise added ``Register payment'') and $\{2, 12\}$ (considering the noise swapped ``Issue invoice'' with ``Register payment'') could be valid.
For this reason, as a measure of goodness to assess if the computed state is ``future-equivalent'' w.r.t.\ the (unknown) real state, we report on whether the state enables the next activity recorded in the ongoing case.\footnote{Given that we measure the goodness of the computed state based on whether it enables the next observed activity, one may suggest that any technique that predicts the next activity in a case is a possible baseline in this evaluation. However, we note that an approach that predicts the next activity (e.g.\ using a deep learning technique), is not a suitable baseline because it does not return a state (i.e.\ a marking) in the process model.}

\begin{table}[t]
    \centering
    \caption{
      Accuracy on synthetic datasets (ratio of ongoing cases for which the next recorded activity is enabled in the computed state) with different levels of noise.
      Shadowed cells highlight the best result for each dataset ($\pm0.01$).
    }
    \label{tab:synthetic-evaluation-results}

    \begin{tabular}{llrrrrr}
                             \toprule
                             &                  & \multicolumn{1}{c}{\textbf{Seq}} & \multicolumn{1}{c}{\textbf{Loop}} & \multicolumn{1}{c}{\textbf{K3}} & \multicolumn{1}{c}{\textbf{K5}} & \multicolumn{1}{c}{\textbf{K10}} \\ \midrule \midrule

    \multirow{5}{*}{\rotatebox[origin=c]{90}{Raw}}     & $\mathit{TokenR}$  & \highcell1.00 & \highcell1.00 & \highcell1.00 & \highcell1.00 & \highcell1.00 \\ 
                                                       & $\mathit{PrefAl}$  & \highcell1.00 & \highcell1.00 & \highcell1.00 & \highcell1.00 & \highcell1.00 \\ 
                                                       & 3-gram             & \highcell1.00 & \highcell1.00 & \highcell1.00 & \medicell0.99 & 0.63          \\ 
                                                       & 5-gram             & \highcell1.00 & \highcell1.00 & \highcell1.00 & \highcell1.00 & 0.88          \\ 
                                                       & 10-gram            & \highcell1.00 & \highcell1.00 & \highcell1.00 & \highcell1.00 & \highcell1.00 \\ \midrule 

    \multirow{5}{*}{\rotatebox[origin=c]{90}{Noise-1}} & $\mathit{TokenR}$  & 0.03          & 0.03          & 0.11          & 0.26          & 0.41          \\ 
                                                       & $\mathit{PrefAl}$  & 0.75          & 0.74          & 0.80          & \highcell0.88 & \highcell0.89 \\ 
                                                       & 3-gram             & \highcell0.82 & \highcell0.81 & \highcell0.83 & 0.82          & 0.60          \\ 
                                                       & 5-gram             & \highcell0.82 & \highcell0.81 & \highcell0.83 & 0.83          & 0.76          \\ 
                                                       & 10-gram            & \highcell0.82 & \highcell0.81 & \highcell0.83 & 0.83          & 0.82          \\ \midrule 

    \multirow{5}{*}{\rotatebox[origin=c]{90}{Noise-2}} & $\mathit{TokenR}$  & 0.04          & 0.04          & 0.04          & 0.08          & 0.17          \\ 
                                                       & $\mathit{PrefAl}$  & 0.58          & 0.57          & \highcell0.68 & \highcell0.75 & \highcell0.79 \\ 
                                                       & 3-gram             & \highcell0.64 & \highcell0.66 & \highcell0.68 & 0.69          & 0.54          \\ 
                                                       & 5-gram             & \highcell0.64 & \highcell0.66 & \highcell0.68 & 0.69          & 0.66          \\ 
                                                       & 10-gram            & \highcell0.64 & \highcell0.66 & \highcell0.67 & 0.70          & 0.70          \\ \midrule 

    \multirow{5}{*}{\rotatebox[origin=c]{90}{Noise-3}} & $\mathit{TokenR}$  & 0.00          & 0.01          & 0.01          & 0.05          & 0.11          \\ 
                                                       & $\mathit{PrefAl}$  & 0.44          & 0.43          & \highcell0.59 & \highcell0.68 & \highcell0.71 \\ 
                                                       & 3-gram             & \highcell0.51 & \highcell0.55 & 0.56          & 0.62          & 0.50          \\ 
                                                       & 5-gram             & \highcell0.51 & \highcell0.55 & 0.56          & 0.61          & 0.59          \\ 
                                                       & 10-gram            & \highcell0.51 & \highcell0.55 & 0.55          & 0.61          & 0.62          \\ \bottomrule 
    \end{tabular}
\end{table}

\medskip
\noindent\textbf{Results \& Discussion.}
\tablename~\ref{tab:synthetic-evaluation-results} depicts, for each dataset, the percentage of ongoing cases in which the evaluated techniques computed a future-equivalent state.
Both $\mathit{TokenR}$ and $\mathit{PrefAl}$ obtain an accuracy of 100\% in the ``Raw'' datasets, as the absence of noise ensures a perfect replay of the trace prefixes.
For the same reason, the $n$-gram index proposals obtain an accuracy of 100\% in the processes with a $K$-complexity lower or equal to each $n$.
As commented in \sectionname~\ref{subsec:n-gram-index}, in fitting traces, the last $K$ executed activities are enough to denote the ongoing state of a case.

\begin{table*}[t]
    \centering
    \caption{
      Characteristic of the (processed) real-life event logs and ongoing cases used in the evaluation of RQ2 and RQ3.
    }
    \label{tab:real-life-logs-characteristics}
    
    \begin{tabular}{l r r r r r r r r r}
                          \toprule
                          & & & & \multicolumn{3}{c}{\textbf{Trace length}} & \multicolumn{3}{c}{\textbf{Ongoing cases length}}      \\ \cmidrule{5-10}
                          & \multicolumn{1}{c}{\multirow{-2}{*}{\# Cases}} & \multicolumn{1}{c}{\multirow{-2}{*}{\# Events}} & \multicolumn{1}{c}{\multirow{-2}{*}{\# Activities}} & \multicolumn{1}{c}{Min} & \multicolumn{1}{c}{Median} & \multicolumn{1}{c}{Max} & \multicolumn{1}{c}{Min} & \multicolumn{1}{c}{Median} & \multicolumn{1}{c}{Max} \\ \midrule \midrule
    BPIC12                & 13,087  & 164,506   & 23 & 3  & 7  & 96    & 1 & 3 & 81 \\
    $\text{BPIC13}_{inc}$ & 7,554   & 65,533    & 4  & 1  & 6  & 123   & 1 & 3 & 110 \\
    BPIC14                & 46,616  & 466,737   & 39 & 1  & 7  & 178   & 1 & 3 & 167 \\
    BPIC17                & 31,509  & 582,374   & 24 & 8  & 18 & 62    & 1 & 8 & 54 \\
    BPIC18                & 43,809  & 2,514,266 & 41 & 24 & 49 & 2,973 & 1 & 26 & 1,579 \\
    BPIC19                & 251,734 & 1,595,923 & 42 & 1  & 5  & 990   & 1 & 2 & 663 \\
    $\text{BPIC20}_{dom}$ & 10,500  & 56,437    & 17 & 1  & 5  & 24    & 1 & 3 & 13 \\
    $\text{BPIC20}_{int}$ & 6,449   & 72,151    & 34 & 3  & 10 & 27    & 1 & 5 & 23 \\
    $\text{BPIC20}_{pre}$ & 2,099   & 18,246    & 29 & 1  & 8  & 21    & 1 & 4 & 17 \\
    $\text{BPIC20}_{req}$ & 6,886   & 36,796    & 19 & 1  & 5  & 20    & 1 & 2 & 13 \\
    $\text{BPIC20}_{tra}$ & 7,065   & 86,581    & 51 & 3  & 11 & 90    & 1 & 5 & 44 \\
    Sepsis                & 1,050   & 15,214    & 16 & 3  & 13 & 185   & 1 & 6 & 85 \\ \bottomrule
    \end{tabular}
    
\end{table*}

\begin{table*}[t]
    \centering
    \caption{
      Fitness of the discovered process models.
    }
    \label{tab:real-life-process-model}
    
    \begin{tabular}{l r r r r r r}
                          \toprule
                          & \multicolumn{2}{c}{\textbf{$\text{IMf}_{10}$}}  & \multicolumn{2}{c}{\textbf{$\text{IMf}_{20}$}}  & \multicolumn{2}{c}{\textbf{$\text{IMf}_{50}$}}      \\ \cmidrule{2-3} \cmidrule{4-5} \cmidrule{6-7}
                          & \multicolumn{1}{c}{\textbf{Fitness}} & \multicolumn{1}{c}{\textbf{\# AND-splits}} & \multicolumn{1}{c}{\textbf{Fitness}} & \multicolumn{1}{c}{\textbf{\# AND-splits}} & \multicolumn{1}{c}{\textbf{Fitness}} & \multicolumn{1}{c}{\textbf{\# AND-splits}} \\ \midrule \midrule
    BPIC12                & 0.94 & 5 & 0.94 & 2 & 0.80 & 3 \\
    $\text{BPIC13}_{inc}$ & 0.99 & 1 & 0.96 & 1 & 0.78 & 1 \\
    BPIC14                & -    & 6 & 0.98 & 1 & 0.99 & 0 \\
    BPIC17                & 0.99 & 0 & 0.93 & 2 & 0.76 & 2 \\
    BPIC18                & -    & 5 & -    & 5 & -    & 2 \\
    BPIC19                & -    & 4 & -    & 5 & -    & 6 \\
    $\text{BPIC20}_{dom}$ & 0.95 & 0 & 0.94 & 0 & 0.94 & 0 \\
    $\text{BPIC20}_{int}$ & 0.95 & 3 & 0.89 & 2 & 0.85 & 0 \\
    $\text{BPIC20}_{pre}$ & 0.94 & 5 & 0.88 & 4 & 0.81 & 1 \\
    $\text{BPIC20}_{req}$ & 0.96 & 3 & 0.92 & 1 & 0.92 & 1 \\
    $\text{BPIC20}_{tra}$ & -    & 6 & 0.76 & 6 & 0.67 & 5 \\
    Sepsis                & 0.98 & 2 & 0.96 & 2 & 0.86 & 2 \\ \bottomrule
    \end{tabular}
    
\end{table*}

In the datasets with noise, as expected, the accuracy of all techniques decreases as the noise level increases.
However, it does not affect all proposals in the same way.
$\mathit{TokenR}$ presents the higher impact, obtaining the lowest accuracy.
This is due to the artificial addition of tokens in unfitting scenarios, which results in unreachable markings.
Regarding the other techniques, in the processes with $K$-complexity of 1, the noise has a stronger (negative) impact in $\mathit{PrefAl}$, leading to our proposal to present better accuracy in all datasets.\footnote{The performance of the evaluated $n$-gram index proposals is similar in ``Seq'', ``Loop'', and ``K3'' because, as the $K$-complexity of these models is lower or equal to 3, an index with $n=3$ is already complete. Similarly, $5$-gram and $10$-gram proposals present identical performance in ``K5''.}
However, this ranking is inverted as the $K$-complexity of the process increases. 
The reasoning behind this phenomenon comes from the fact that the $n$-gram index proposal estimates the ongoing state based on, at most, the last $n$ executed activities, while $\mathit{PrefAl}$ uses the entire ongoing case to find the best alignment.
The smaller the $K$-complexity of the process -- thus, the smaller the $n$ needed --, the lower the probability of potential noise affecting the $n$-gram index proposal.
At the same time, when affecting both techniques, the noise has a lower impact in $\mathit{PrefAl}$.
This is due to our proposal performing an exact matching for the $n$-gram prefix, while $\mathit{PrefAl}$ allows for deviations when aligning the ongoing case with the model.


In conclusion, for fitting traces, $\mathit{PrefAl}$ always computes the correct state.
As expected, the $n$-gram index approach fails when $n \le$ the $K$-complexity of the model.

In the presence of noise, the $n$-gram approach achieves higher accuracy for models with low $K$-complexity.
However, when the $K$-complexity of the model increases, it underperforms the $\mathit{PrefAl}$. 
This result is expected. 
When the $K$-complexity is high, the $n$-gram approach does not have sufficient information to compute the correct state.
The presence of noise aggravates this lack of information.
Meanwhile, when the $K$-complexity is low, the $n$-gram approach has sufficient information to compute the correct state, while being less likely to be affected by the noise since it only looks at the last $n$ events in the prefix.
The $\mathit{PrefAl}$ is more likely to be affected by noise, since it looks at the whole prefix.
Prefix alignment techniques are designed to find a minimal set of ``skip event'' (move-on-log) or ``add event'' (move-on-model) operations to transform a given trace prefix into a prefix that can be parsed by a process model.
In the presence of noise, these techniques find an alignment containing a minimal set of operations to correct this noise, but when this alignment is replayed against the model, the replay may or may not lead to the state that would be reached had the noise not been present.

Regarding $\mathit{TokenR}$, this approach often returns unreachable states in the presence of noise.
The results suggest that $\mathit{TokenR}$ is not a suitable approach for state computation when the input trace prefixes do not fit the behavior of the model.
Furthermore, in order to evaluate if $\mathit{TokenR}$ can produce unreachable states even for fitting traces, we ran all techniques for a fitting event log of the process model depicted in \figurename~\ref{subfig:ambiguous-model}.
Results showed that $\mathit{TokenR}$ produced unreachable markings in 2\% of the cases, while both $\mathit{PrefAl}$ and our proposal produced, as expected, a reachable marking in all cases.


\subsection{Real-life Evaluation}

This section describes the accuracy evaluation performed on real-life scenarios (EQ2), as well as the evaluation of the efficiency of our proposal (EQ3).

\medskip
\noindent\textbf{Datasets \& Setup.}
We selected 12 real-life event logs from the publicly available collection at the 4TU Centre for Research Data\footnote{\url{https://data.4tu.nl/}} corresponding to business processes, as opposed to, e.g., software development processes.
We discarded the logs of the Business Process Intelligence Challenges (BPIC) of 2011 and 2015, as the high number of activities (over 300) and low variant frequency prevented all approaches from retrieving a result in a reasonable time.
We also discarded the Road Traffic Fines and BPIC 2013 (closed and open problems) due to having an average case length below 4, which would lead to most ongoing cases having only 1 or 2 events.
The remaining selected logs correspond to the BPIC of 2012, 2013 (incidents), 2014, 2017, 2018, 2019, and 2020, and a Sepsis patient treatment process.
We preprocessed the logs with lifecycle information by retaining only the events corresponding to the activity completion, i.e., complete, closed, etc.
Then, we obtained the ongoing cases by retaining the first $m$ recorded activities of each complete case, being $m$ a random number between one and the number of events recorded in each case minus one.
\tablename~\ref{tab:real-life-logs-characteristics} depicts the characteristics of these preprocessed logs, containing from 4 to 51 activities and from 15,000 to 2,500,000 events.
Regarding the ongoing cases, their minimum length is 1, their median ranges from 2 to 26, and their maximum from 13 to 1,579.

\begin{table}[t]
    \centering
    \caption{
      Accuracy on real-life logs (ratio of ongoing cases for which the next recorded activity is enabled in the computed state).
      Shadowed cells highlight the best result for each dataset ($\pm0.01$).
    }
    \label{tab:real-life-evaluation-results}
 
    \begin{tabular}{l l r r r r}
                               \toprule
                               &              & \multicolumn{1}{c}{$\mathit{PrefAl}$} & \multicolumn{1}{c}{\textbf{3-gram}} & \multicolumn{1}{c}{\textbf{4-gram}} & \multicolumn{1}{c}{\textbf{5-gram}} \\ \midrule \midrule

    \multirow{12}{*}{\rotatebox[origin=c]{90}{$\text{IMf}_{10}$}} & BPIC12                  & \highcell0.97 & 0.94          & 0.95          & \medicell0.96 \\ 
                                                                  & $\text{BPIC13}_{inc}$   & \highcell0.99 & 0.97          & 0.97          & 0.97          \\ 
                                                                  & BPIC14                  & -             & \highcell0.94 & \highcell0.94 & \highcell0.94 \\  
                                                                  & BPIC17                  & \medicell0.98 & \highcell0.99 & \highcell0.99 & \highcell0.99 \\ 
                                                                  & BPIC18                  & -             & -             & -             & -             \\  
                                                                  & BPIC19                  & -             & \highcell0.81 & \highcell0.81 & -             \\  
                                                                  & $\text{BPIC20}_{dom}$   & 0.92          & \highcell0.94 & \highcell0.94 & \highcell0.94 \\  
                                                                  & $\text{BPIC20}_{int}$   & \highcell0.92 & \medicell0.91 & 0.90          & \medicell0.91 \\  
                                                                  & $\text{BPIC20}_{pre}$   & \medicell0.91 & \highcell0.92 & \highcell0.92 & \highcell0.92 \\  
                                                                  & $\text{BPIC20}_{req}$   & 0.95          & \highcell0.97 & \highcell0.97 & \highcell0.97 \\  
                                                                  & $\text{BPIC20}_{tra}$   & -             & 0.84          & \highcell0.86 & \highcell0.86 \\  
                                                                  & Sepsis                  & \highcell0.98 & 0.81          & 0.86          & 0.85          \\ \midrule 
                                                                  
    \multirow{12}{*}{\rotatebox[origin=c]{90}{$\text{IMf}_{20}$}} & BPIC12                  & \highcell0.96 & \medicell0.95 & \highcell0.96 & \highcell0.96 \\ 
                                                                  & $\text{BPIC13}_{inc}$   & 0.78          & \highcell0.91 & \highcell0.91 & \highcell0.91 \\ 
                                                                  & BPIC14                  & \highcell0.97 & \highcell0.97 & \highcell0.97 & \highcell0.97 \\  
                                                                  & BPIC17                  & \highcell0.93 & 0.91          & 0.91          & 0.91          \\  
                                                                  & BPIC18                  & -             & \medicell0.89 & \highcell0.90 & \highcell0.90 \\  
                                                                  & BPIC19                  & -             & \medicell0.73 & \highcell0.74 & -             \\  
                                                                  & $\text{BPIC20}_{dom}$   & 0.82          & \highcell0.89 & \highcell0.89 & \highcell0.89 \\  
                                                                  & $\text{BPIC20}_{int}$   & 0.80          & \highcell0.86 & \highcell0.86 & \highcell0.86 \\  
                                                                  & $\text{BPIC20}_{pre}$   & 0.80          & \highcell0.89 & \highcell0.89 & \highcell0.89 \\  
                                                                  & $\text{BPIC20}_{req}$   & 0.83          & \highcell0.90 & \highcell0.90 & \highcell0.90 \\  
                                                                  & $\text{BPIC20}_{tra}$   & 0.66          & \highcell0.80 & \medicell0.79 & \highcell0.80 \\  
                                                                  & Sepsis                  & \highcell0.99 & 0.77          & 0.78          & 0.81          \\ \midrule 
                                                                  
    \multirow{12}{*}{\rotatebox[origin=c]{90}{$\text{IMf}_{50}$}} & BPIC12                  & 0.59          & \highcell0.61 & \highcell0.61 & \highcell0.61 \\  
                                                                  & $\text{BPIC13}_{inc}$   & 0.62          & \medicell0.71 & \highcell0.72 & \highcell0.72 \\ 
                                                                  & BPIC14                  & \highcell1.00 & \highcell1.00 & \highcell1.00 & \highcell1.00 \\ 
                                                                  & BPIC17                  & 0.68          & \highcell0.72 & \highcell0.72 & \highcell0.72 \\ 
                                                                  & BPIC18                  & -             & \highcell0.83 & \highcell0.83 & \highcell0.83 \\ 
                                                                  & BPIC19                  & -             & \medicell0.74 & \highcell0.75 & \highcell0.75 \\ 
                                                                  & $\text{BPIC20}_{dom}$   & 0.82          & \highcell0.89 & \highcell0.89 & \highcell0.89 \\ 
                                                                  & $\text{BPIC20}_{int}$   & 0.77          & \highcell0.86 & \highcell0.86 & \highcell0.86 \\ 
                                                                  & $\text{BPIC20}_{pre}$   & 0.74          & \highcell0.86 & \highcell0.86 & \highcell0.86 \\ 
                                                                  & $\text{BPIC20}_{req}$   & 0.83          & \highcell0.90 & \highcell0.90 & \highcell0.90 \\ 
                                                                  & $\text{BPIC20}_{tra}$   & \highcell0.66 & 0.63          & 0.63          & \medicell0.65 \\ 
                                                                  & Sepsis                  & \highcell0.91 & 0.73          & 0.74          & 0.76          \\ \bottomrule 
    \end{tabular}
\end{table}

For each log, we used the Inductive Miner infrequent~\cite{DBLP:conf/bpm/LeemansFA13} to discover a WF-net.
To analyze how the accuracy varies depending on the fitness, we discovered three process models (with thresholds of 10\%, 20\%, and 50\%) per log.
\tablename~\ref{tab:real-life-process-model} depicts the characteristics of the discovered models.
The fitness ranges from 0.94 to 0.99 (10\% threshold), from 0.88 to 0.98 (20\% threshold), and from 0.67 to 0.99 (50\% threshold).
Six out of the 36 models are perfectly sequential, while the rest contain from 1 to 6 AND-split structures.

Due to the high tendency of $\mathit{TokenR}$ to create unreachable markings with additional tokens (see evaluation of EQ1), we discarded this approach for the evaluation of EQ2 and EQ3.
Accordingly, this evaluation compares the performance and efficiency of $\mathit{PrefixAl}$ and multiple versions of our proposal.

Due to the lack of ground truth in the case of real-life processes, as the state of each ongoing case is unknown, the evaluation of EQ2 is built upon the same measure of goodness as EQ1.

\begin{table*}[t]
    \centering
    \caption{
      Runtimes on the real-life logs (in seconds needed to handle one ongoing case).
      The shadowed cells denote the lowest runtimes per dataset ($\pm0.01$). Index construction times are also included.
    }
    \label{tab:real-life-evaluation-runtimes}

    \begin{tabular}{l l r r r r r r r r}
                               \toprule
                               &              &  & \multicolumn{7}{c}{\textbf{N-gram index}}                                                                                                                                                                                                                                                                                                                                                                                                    \\ \cmidrule{4-10}
                               &              & \multicolumn{1}{c}{\multirow{-2}{*}{$\mathit{PrefAl}$}}  & \multicolumn{1}{c}{\textbf{Reachability}} & \multicolumn{2}{c}{\textbf{3-gram}}                                                      & \multicolumn{2}{c}{\textbf{4-gram}}                                                      & \multicolumn{2}{c}{\textbf{5-gram}}    \\ \cmidrule{3-3} \cmidrule{5-6} \cmidrule{7-8} \cmidrule{9-10}
                               &              & \multicolumn{1}{c}{\textbf{Avg}} & \multicolumn{1}{c}{\multirow{-2}{*}{\textbf{Graph}}}  & \multicolumn{1}{c}{\textbf{Index}} & \multicolumn{1}{c}{\textbf{Avg}} & \multicolumn{1}{c}{\textbf{Index}} & \multicolumn{1}{c}{\textbf{Avg}} & \multicolumn{1}{c}{\textbf{Index}} & \multicolumn{1}{c}{\textbf{Avg}} \\ \midrule \midrule
                               
    \multirow{12}{*}{\rotatebox[origin=c]{90}{$\text{IMf}_{10}$}} & BPIC12                  & 0.50 & 0.05  & 0.01 & \highcell4e-6 & 0.04  & \highcell3e-6 & 0.18   & \highcell5e-6 \\ 
                                                                  & $\text{BPIC13}_{inc}$   & 0.17 & 0.01  & 0.01 & \highcell5e-6 & 0.01  & \highcell5e-6 & 0.01   & \highcell4e-6 \\ 
                                                                  & BPIC14                  & -    & 13.15 & 0.43 & \highcell4e-6 & 4.07  & \highcell3e-6 & 65.06  & \highcell3e-6 \\ 
                                                                  & BPIC17                  & 0.10 & 0.01  & 0.01 & \highcell4e-6 & 0.01  & \highcell2e-6 & 0.01   & \highcell3e-6 \\ 
                                                                  & BPIC18                  & -    & -     & -    & -             & -     & -             & -      & -             \\ 
                                                                  & BPIC19                  & -    & 6.42  & 1.77 & \highcell1e-5 & 41.66 & \highcell1e-5 & -      & -             \\
                                                                  & $\text{BPIC20}_{dom}$   & 0.03 & 0.01  & 0.01 & \highcell2e-6 & 0.01  & \highcell3e-6 & 0.01   & \highcell2e-6 \\ 
                                                                  & $\text{BPIC20}_{int}$   & 0.25 & 0.03  & 0.01 & \highcell5e-6 & 0.01  & \highcell5e-6 & 0.02   & \highcell5e-6 \\ 
                                                                  & $\text{BPIC20}_{pre}$   & 0.39 & 0.03  & 0.01 & \highcell2e-6 & 0.01  & \highcell2e-6 & 0.02   & \highcell2e-6 \\ 
                                                                  & $\text{BPIC20}_{req}$   & 0.02 & 0.01  & 0.01 & \highcell2e-6 & 0.01  & \highcell5e-6 & 0.01   & \highcell4e-6 \\ 
                                                                  & $\text{BPIC20}_{tra}$   & -    & 71.56 & 3.19 & \highcell4e-7 & 19.37 & \highcell4e-7 & 135.69 & \highcell4e-7 \\ 
                                                                  & Sepsis                  & 0.43 & 0.89  & 0.06 & \highcell5e-6 & 0.35  & \highcell2e-5 & 1.90   & \highcell2e-5 \\ \midrule 
                                                                  
    \multirow{12}{*}{\rotatebox[origin=c]{90}{$\text{IMf}_{20}$}} & BPIC12                  & 0.31 & 0.01  & 0.01 & \highcell3e-6 & 0.01  & \highcell3e-6 & 0.02  & \highcell1e-6 \\ 
                                                                  & $\text{BPIC13}_{inc}$   & 0.14 & 0.01  & 0.01 & \highcell6e-6 & 0.01  & \highcell4e-6 & 0.01  & \highcell4e-6 \\ 
                                                                  & BPIC14                  & 0.79 & 0.01  & 0.01 & \highcell3e-6 & 0.01  & \highcell2e-6 & 0.01  & \highcell3e-6 \\ 
                                                                  & BPIC17                  & 0.48 & 0.05  & 0.01 & \highcell6e-6 & 0.02  & \highcell5e-6 & 0.06  & \highcell7e-6 \\ 
                                                                  & BPIC18                  & -    & 0.41  & 0.06 & \highcell1e-5 & 0.42  & \highcell1e-5 & 3.29  & \highcell1e-5 \\ 
                                                                  & BPIC19                  & -    & 70.60 & 3.74 & \highcell2e-5 & 65.75 & \highcell2e-5 & -     & -             \\ 
                                                                  & $\text{BPIC20}_{dom}$   & 0.02 & 0.01  & 0.01 & \highcell2e-6 & 0.01  & \highcell2e-6 & 0.01  & \highcell2e-6 \\ 
                                                                  & $\text{BPIC20}_{int}$   & 0.18 & 0.01  & 0.01 & \highcell2e-6 & 0.01  & \highcell4e-6 & 0.01  & \highcell3e-6 \\ 
                                                                  & $\text{BPIC20}_{pre}$   & 0.19 & 0.02  & 0.01 & \highcell5e-6 & 0.01  & \highcell2e-6 & 0.01  & \highcell1e-5 \\ 
                                                                  & $\text{BPIC20}_{req}$   & 0.02 & 0.01  & 0.01 & \highcell1e-6 & 0.01  & \highcell3e-6 & 0.01  & \highcell1e-6 \\ 
                                                                  & $\text{BPIC20}_{tra}$   & 3.00 & 0.38  & 0.13 & \highcell5e-6 & 1.95  & \highcell5e-6 & 27.48 & \highcell6e-6 \\ 
                                                                  & Sepsis                  & 0.30 & 1.04  & 0.04 & \highcell9e-6 & 0.22  & \highcell2e-5 & 1.31  & \highcell1e-5 \\ \midrule 
                                                                  
    \multirow{12}{*}{\rotatebox[origin=c]{90}{$\text{IMf}_{50}$}} & BPIC12                  & 0.18 & 0.01 & 0.01 & \highcell4e-6 & 0.01  & \highcell4e-6 & 0.01   & \highcell3e-6 \\ 
                                                                  & $\text{BPIC13}_{inc}$   & 0.15 & 0.01 & 0.01 & \highcell3e-6 & 0.01  & \highcell3e-6 & 0.01   & \highcell3e-6 \\ 
                                                                  & BPIC14                  & 1.31 & 0.03 & 0.01 & \highcell2e-6 & 0.01  & \highcell2e-6 & 0.01   & \highcell3e-6 \\ 
                                                                  & BPIC17                  & 0.39 & 0.02 & 0.01 & \highcell5e-6 & 0.02  & \highcell6e-6 & 0.06   & \highcell7e-6 \\ 
                                                                  & BPIC18                  & -    & 0.74 & 0.82 & \highcell1e-5 & 18.98 & \highcell1e-5 & 439.16 & \highcell1e-5 \\ 
                                                                  & BPIC19                  & -    & 8.46 & 2.26 & \highcell1e-5 & 26.93 & \highcell1e-5 & 555.44 & \highcell1e-5 \\ 
                                                                  & $\text{BPIC20}_{dom}$   & 0.02 & 0.01 & 0.01 & \highcell2e-6 & 0.01  & \highcell1e-6 & 0.01   & \highcell4e-6 \\ 
                                                                  & $\text{BPIC20}_{int}$   & 0.14 & 0.01 & 0.01 & \highcell5e-6 & 0.01  & \highcell2e-6 & 0.01   & \highcell1e-6 \\ 
                                                                  & $\text{BPIC20}_{pre}$   & 0.10 & 0.01 & 0.01 & \highcell1e-5 & 0.01  & \highcell2e-6 & 0.01   & \highcell2e-7 \\ 
                                                                  & $\text{BPIC20}_{req}$   & 0.02 & 0.01 & 0.01 & \highcell4e-6 & 0.01  & \highcell2e-6 & 0.01   & \highcell2e-6 \\ 
                                                                  & $\text{BPIC20}_{tra}$   & 1.40 & 0.11 & 0.03 & \highcell6e-6 & 0.16  & \highcell4e-6 & 0.99   & \highcell6e-6 \\ 
                                                                  & Sepsis                  & 0.19 & 0.10 & 0.05 & \highcell1e-5 & 0.26  & \highcell1e-5 & 1.28   & \highcell2e-5 \\ \bottomrule 
                                                                  
    \end{tabular}
\end{table*}

\smallskip
\noindent\textbf{Results \& Discussion.}
Regarding EQ2, \tablename~\ref{tab:real-life-evaluation-results} shows the accuracy of the evaluated methods.
The $n$-gram index yields higher accuracy than the $\mathit{PrefAl}$ technique in most cases.
Although the differences are negligible in the models with higher complexity ($\text{IMf}_{10}$), they grow as the support of the model decreases.
Furthermore, in 7 out of the 36 scenarios, $\mathit{PrefAl}$ was not able to compute the state of ongoing cases in a reasonable time, while the $n$-gram index method did.
The Sepsis Cases dataset is an exception, where $\mathit{PrefAl}$ outperforms the $n$-gram index method.
As expected, the accuracy of the techniques generally decreases with the support.

Regarding EQ3, \tablename~\ref{tab:real-life-evaluation-runtimes} shows the runtimes of each proposal.
$\mathit{PrefAl}$ reuses pre-computed heuristics to reduce the space explored by the $A^{*}$ which, in logs with tens of thousands of cases, reduces the average runtime of the first cases.
In this paper, we study the applicability of the proposals for scenarios where the expected number of cases to process simultaneously amounts to a few thousand.
Accordingly, we present the runtime of $\mathit{PrefAl}$ as the average of the first 1,000 processed cases.
For the $n$-gram index proposal, we report separately the creation of the reachability graph and $n$-gram index (offline phase) and the average time required to process one ongoing case (online phase).

In the online phase, our proposal achieves a throughput of around 100~000 cases per second, while $\mathit{PrefAl}$ ranges from 0.33 to 50 cases per second, with an average of nearly 2 cases per second.
Furthermore, in BPIC18 and BPIC19, we stopped the execution of $\mathit{PrefAl}$ after 2 hours as it had processed less than 20 cases.
Regarding the offline phase, the reachability graph is built in less than 1 second in most cases, and less than 2 minutes in the worst case.
The $n$-gram indexes are generally computed in less than 5 seconds, with almost 10 minutes in the worst case. 

\subsection{Limitations and Validity}

This section introduces the limitations of the proposed approach, as well as the threats to validity associated with the empirical evaluation.

\smallskip
\noindent\textbf{Limitations.}
A limitation of our proposal is the restriction to sound WF-net, instead of arbitrary Petri nets.
However, as stated in \sectionname~\ref{sec:background}, sound WF-nets ensure desirable properties for modeling business processes such as \textit{i)} the absence of deadlocks; \textit{ii)} ensuring that, at any point in the execution of a process instance, there is an executable path to terminate the case; and \textit{iii} the absence of activities that can never be executed~\cite{iordache2006deadlock,DBLP:books/sp/Aalst16}.

Notwithstanding the above, we foresee that the proposed approach can be applied to unsound WF-nets by: \textit{i)} representing a marking with a multiset instead of a set, and \textit{ii)} limiting the number of tokens in each place when computing the reachability graph in order to avoid infinite state spaces.

Regarding the presence of non-observed activities in the ongoing cases, our proposal is to, prior to the search, discard from the trace prefix all activities that are not present in the model.
This decision is equivalent to that of performing a ``\textit{move-on-log}'' when aligning a case with a model.
If there is no evidence of such activity in the process model, the safer decision is to assume that no movements were made in the state of the ongoing case.

\smallskip
\noindent\textbf{Threats to validity.}
The reported evaluation is potentially affected by the following threats to validity.
First, regarding \textit{external validity}, the experiments rely only on five simulated and twelve real-life processes.
The results could be different for other datasets.
We have mitigated this threat by selecting datasets from processes across different domains.
Second, regarding \textit{construct validity}, in the evaluation of EQ1 and EQ2, we used a measure of goodness based on the enablement of the next recorded activity.
The results could be different with other measures, e.g., considering the ``\textit{replayability}'' of the remaining case rather than only the next activity.

\section{Conclusion and Future Work\label{sec:conclusions}}

We presented a method to efficiently compute the state of an ongoing case w.r.t.\ a process model represented as a workflow net.
Given a process model, the method generates a reachability graph and builds an $n$-gram index that associates each sequence of $n$ (or fewer) consecutive activities generated by the model, with the state(s) in the reachability graph that this $n$-gram leads to.
This $n$-gram index is computed offline and stored as a hash table.
At runtime, the state of an ongoing case, given its trace prefix, is computed in linear time on parameter $n$ (irrespective of the length of the ongoing case) by searching for the last $n$ activities of the prefix, or the last $m < n$ activities, if this search fails.


The synthetic evaluation showed that our proposal is less likely to be affected by noise than the $\mathit{PrefAl}$ baseline, as it only works with the last $n$ executed activities.
However, our proposal may fail to compute the correct state in complex processes.
On the real-life logs, our method achieves a throughput of hundreds of thousands of cases per second, with an accuracy comparable to or above the baseline, w.r.t.\ its ability to correctly predict that the next activity is enabled in the returned state.
These results hint that, in real-life processes, the next activities of an ongoing case depend more on the recently observed activities than on activities at the start of a case.

We also studied the problem of computing a reachability graph of a process model considering lazy vs. eager policies for silent transition traversal.
We proposed an algorithm to compute a reachability graph that allows for the replay of fitting traces in linear time. 

When the ending $n$-gram of a trace prefix cannot be found, we iteratively search for smaller $m$-grams $(m < n)$.
This is, in essence, an approach to finding a partial $n$-gram match.
In future work, we plan to experiment with partial $n$-gram matching techniques to retrieve the closest matching $n$-gram(s).
A partial matching approach could also help us to reduce the number of $n$-grams we need to index.
Searching in the index with small variations of this $n$-gram may lead to a better result when the $n$-gram does not have a match itself, or reduce the number of states when the $n$-gram itself returns a big number of them.
We also plan to study the possibility of storing the reachability graph in a graph database, and query directly for the markings that are at the end of an arc sequence with the $n$-gram as labels.
This would sacrifice runtime, but avoid the need to compute and store the $n$-gram index.

\smallskip
\noindent\textbf{Reproducibility}
The proposed method has been implemented as a Python package installable from pip (\href{https://pypi.org/project/ongoing-process-state/}{\texttt{ongoing-process-state}}), and the code and scripts to reproduce the experiments are available at: {\url{https://github.com/AutomatedProcessImprovement/ongoing-process-state/tree/IEEETSC-2}}.
The datasets and evaluation results are available at: {\url{https://doi.org/10.5281/zenodo.14679920}}.

\section*{Acknowledgements}

This research is funded by the Estonian Research Council (PRG1226).

    \bibliographystyle{IEEEtran}
    \bibliography{references}

\vspace*{-10mm}
    \begin{IEEEbiography}[{\includegraphics[width=0.9in,height=1.2in,clip,keepaspectratio]{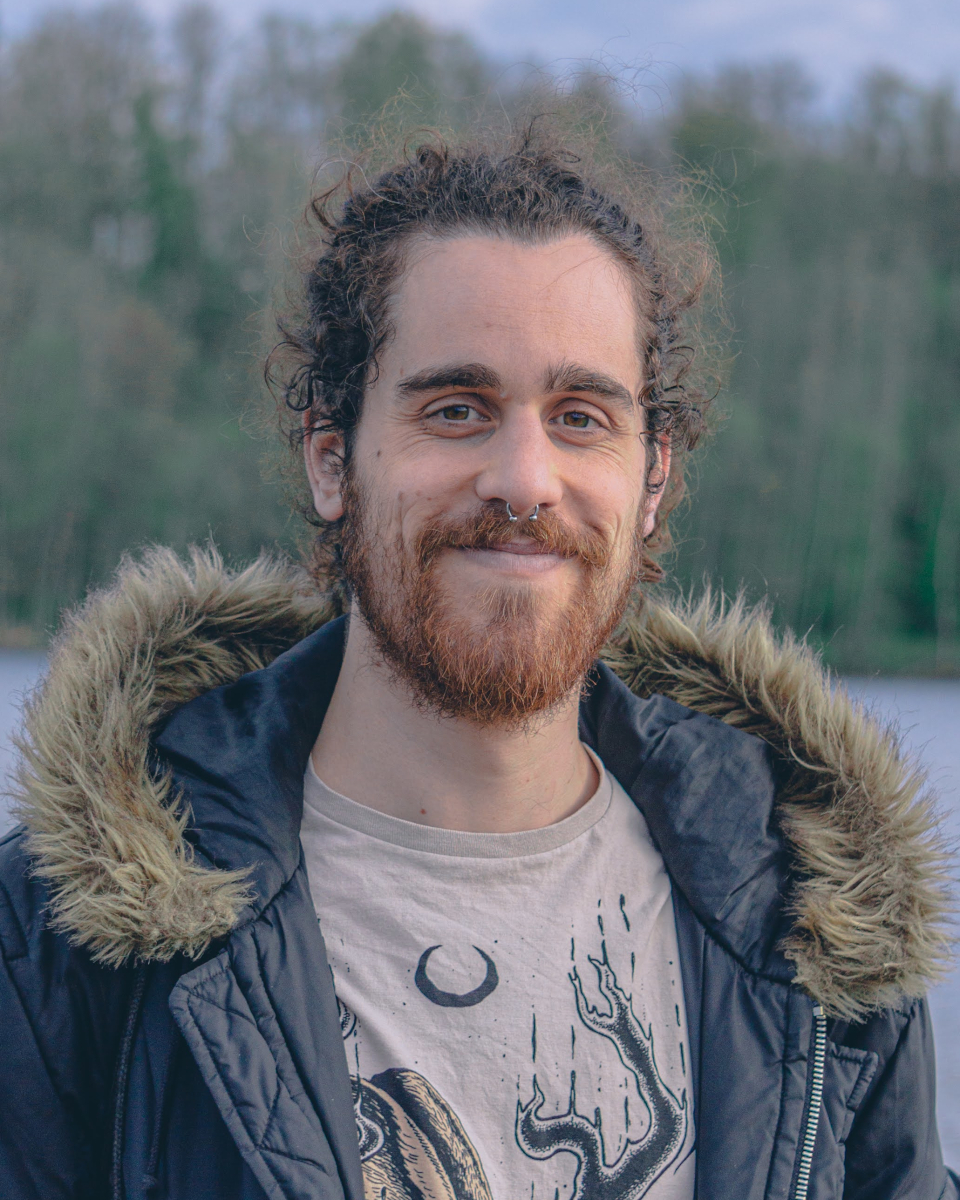}}]{David Chapela-Campa}
    obtained his PhD in Computer Science from the University of Santiago de Compostela in 2021. 
    He is an Assistant Professor of Information Systems at the University of Tartu.
    His research focuses on process mining and business process management, specifically on data-driven techniques for business process simulation and optimization.
    \end{IEEEbiography}
    \vspace*{-11mm}

    \begin{IEEEbiography}[{\includegraphics[width=0.9in,height=1.2in,clip,keepaspectratio]{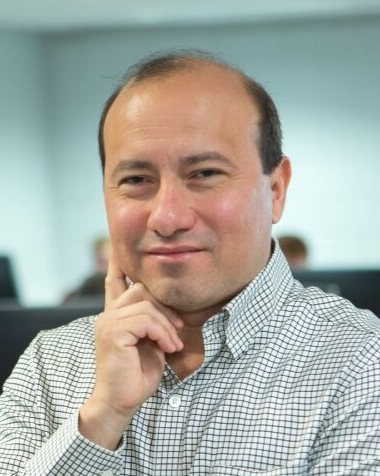}}]{Marlon Dumas}
    is a Professor of Information Systems at the University of Tartu, and co-founder of Apromore -- a company dedicated to developing and commercializing open-source process mining solutions.
    His research focuses on data-driven methods for business process management, including process mining and predictive process monitoring.
    \end{IEEEbiography}

\end{document}